\documentclass[12pt,preprint]{aastex}
\usepackage{amsmath}

\def\edcomment#1{\iffalse\marginpar{\raggedright\sl#1\/}\else\relax\fi}
\reversemarginpar

\def\p{\partial}

\def\nab{\mbox{\boldmath $\nabla$}}
\def\rb{\bar{\rho}}
\def\tb{\bar{T}}
\def\sb{\bar{S}}

\def\vrr{\tilde{v}_r} 
\def\vtr{\tilde{v}_{\theta}}
\def\vphr{\tilde{v}_{\phi}}
\def\vvr{\tilde{v}}
\def\brr{\tilde{B}_r} 
\def\btr{\tilde{B}_{\theta}}
\def\bphr{\tilde{B}_{\phi}}
\def\bbr{\tilde{B}}
\def\EE{\cal E}

\newcommand{\pd}{\partial}
\newcommand{\rh}{\bar{\rho}}
\newcommand{\Sh}{\bar{S}}
\newcommand{\Th}{\bar{T}}
\newcommand{\ph}{\bar{P}}
\newcommand{\degrees}{^\circ}

\newcommand{\BB}{{\bf B}}
\newcommand{\JJ}{{\bf J}}
\newcommand{\curl}{\mbox{\boldmath $\nabla \times$}}
\newcommand{\cross}{\mbox{\boldmath $\times$}}
\newcommand{\vort}{\mbox{\boldmath $\omega$}}
\newcommand{\uvr}{\mbox{\boldmath $\hat{\bf e}_r$}}
\newcommand{\uvt}{\mbox{\boldmath $\hat{\bf e}_\theta$}}
\newcommand{\uvp}{\mbox{\boldmath $\hat{\bf e}_\phi$}}

\begin{document}

\title{Global-Scale Turbulent Convection and Magnetic Dynamo Action in the 
Solar Envelope}
\author{Allan Sacha Brun}
\affil{DSM/DAPNIA/SAp, CEA Saclay, Gif sur Yvette, 91191 Cedex, France}
\author{Mark S.\ Miesch}
\affil{HAO, NCAR, Boulder, CO 80307-3000}
\author{Juri Toomre}
\affil{JILA, University of Colorado, Boulder, CO 80309-0440}
\begin{abstract}

The operation of the solar global dynamo appears to involve many
dynamical elements, including the generation of fields by the intense
turbulence of the deep convection zone, the transport of these fields
into the tachocline region near the base of the convection zone, the
storage and amplification of toroidal fields in the tachocline by
differential rotation, and the destablization and emergence of such
fields due to magnetic buoyancy.  Self-consistent magnetohydrodynamic
(MHD) simulations which realistically incorporate all of these
processes are not yet computationally feasible, though some elements
can now be studied with reasonable fidelity.  Here we consider the
manner in which turbulent compressible convection within the bulk of
the solar convection zone can generate large-scale magnetic fields
through dynamo action.  We accomplish this through a series of
three-dimensional numerical simulations of MHD convection within
rotating spherical shells using our anelastic spherical harmonic (ASH)
code on massively parallel supercomputers.  Since differential
rotation is a key ingredient in all dynamo models, we also examine
here the nature of the rotation profiles that can be sustained within
the deep convection zone as strong magnetic fields are built and
maintained.  We find that the convection is able to maintain a
solar-like angular velocity profile despite the influence of Maxwell
stresses which tend to oppose Reynolds stresses and thus reduce the
latitudinal angular velocity contrast throughout the convection zone.
The dynamo-generated magnetic fields exhibit a complex structure and
evolution, with radial fields concentrated in downflow lanes and
toroidal fields organized into twisted ribbons which are extended in
longitude and which achieve field strengths of up to 5000 G.  The
flows and fields exhibit substantial kinetic and magnetic helicity
although systematic hemispherical patterns are only apparent in the
former.  Fluctuating fields dominate the magnetic energy and account
for most of the back-reaction on the flow via Lorentz forces.  Mean
fields are relatively weak and do not exhibit systematic latitudinal
propagation or periodic polarity reversals as in the sun.  This may be
attributed to the absence of a tachocline, i.e. a penetrative boundary
layer between the convection zone and the deeper radiative interior
possessing strong rotational shear.  The influence of such a layer
will await subsequent studies.


\end{abstract}

\section{Turbulent Magnetic Sun}

The sun is a magnetic star whose variable activity has profound effects
on our technological society on Earth.  The high speed solar wind and
its energetic particles, coronal mass ejections, and explosive flares
are all linked to the changing magnetic fields within the extended
solar atmosphere.  Such events can serve to damage satellites in space
and power grids on the ground, and interrupt communications.  Thus
there is keen interest in being able to forecast the behavior of the
magnetic structures.  Yet this has proved to be difficult, since the
eruption of new magnetic flux through the solar surface appears to have
a dominant role in the evolution of field configurations in the solar
atmosphere, as does the shuffling of field footpoints by the subsurface
turbulence.

The origin of the solar magnetic fields must rest with dynamo processes
occurring deep within the star in the spherical shell of intensely
turbulent convection that occupies the outer 29\% in radius below the
solar surface.  Within this convection zone, complex interactions
between compressible turbulence and rotation of the star serve to
redistribute angular momentum so that a strong differential rotation is
achieved.  Further, since the fluid is electrically conducting,
currents will flow and magnetic fields must be built.  Yet there are
many fundamental puzzles about the dynamo action that yields the
observed fields.

The magnetic fields, like the underlying turbulence, can be both
orderly on some scales and chaotic on others.  Most striking is that
the sun exhibits 22-year cycles of global magnetic activity, involving
sunspot eruptions with very well defined rules for field parity and
emergence latitudes as the cycle evolves.  Coexisting with these
large-scale ordered magnetic structures are small-scale but intense
magnetic fluctuations that emerge over much of the solar surface, with
little regard for the solar cycle.  This diverse range of activity is
most likely generated by two conceptually distinct magnetic dynamos
(e.g., Weiss 1994; Childress \& Gilbert 1995; Cattaneo 1999; Cattaneo
\& Hughes 2001; Ossendrijver 2003).  These involve a {\sl small-scale
dynamo}, functioning within the intense turbulence of the upper
convection zone, that builds the chaotic magnetic fluctuations, and a
{\sl global dynamo}, operating both within the deeper convection zone
and the strong rotational shear of the tachocline at its base, 
that builds the more ordered fields.

\subsection{Building Magnetic Fields in the Sun}

The pairing of opposite polarity sunspots in the east-west direction
within active regions is most readily interpreted as the surface
emergence of large-scale toroidal field structures.  These structures
are created somewhere below the photosphere and rise upwards, bending
to pierce the photosphere in the form of curved tubes.  The current
paradigm for large-scale dynamo action (e.g., Parker 1993) involves
two major components.  First, strong toroidal field structures must be
generated.  This is believed to occur due to the stretching that any
differential rotation in latitude or radius will impose on any weak
existing poloidal field.  This first process is often referred to as
the $\omega$-effect after its parameterization within the framework of
mean-field electrodynamics (e.g., Moffatt 1978; Krause \& R\"adler
1980; Parker 1989).  Helioseismology has shown that gradients in
angular velocity are particularly strong in the tachocline, pointing
to this interface region between the convection zone and the deeper
radiative interior as the likely site for the generation of strong
toroidal fields.  Second, an inverse process is required to
complete the cycle, regenerating the poloidal field from the toroidal
field.  Different theories exist for the operation of this process
(known as the $\alpha$-effect).  Some have the poloidal field
regenerated at the surface through the breakup and reconnection of the
large-scale field that emerges as active regions, where this field has
gained a poloidal component due to Coriolis forces during its rise,
with meridional flows having a key role in transporting such flux both
poleward and down toward the tachocline (e.g., Babcock 1961; Leighton
1969; Wang \& Sheeley 1991; Durney 1997; Dikpati \& Charbonneau 1999).
Others believe that the poloidal field is regenerated by the
cumulative action of many small-scale cyclonic turbulent motions on
the field throughout the convection zone, rather than just close to
the surface (e.g., Parker 1993).  In either scenario, there is
separation in the sites of generation of toroidal field (in the strong
shear of the tachocline) and regeneration of poloidal field (either
near the surface or in the bulk of the convection zone), yielding what
is now broadly called an {\sl interface dynamo} (Parker 1993).  Recent
mean-field dynamo approaches (e.g., R\"udiger \& Brandenburg 1995;
Tobias 1996, Charbonneau \& MacGregor 1997, Beer et al. 1998) suggest
that such an interface model can circumvent the problem of strong
$\alpha$-quenching by mean magnetic fields (Cattaneo \& Hughes 1996),
thereby being capable of yielding field strengths comparable to those
inferred from observations.

The interface dynamo paradigm is thus based on the following underlying
processes or building blocks:~~{$(a)$} {The $\alpha$-effect:}  the
generation of the background weak poloidal field, either by cyclonic
turbulence within the convection zone or by breakup of active
regions.~~{$(b)$} {The $\beta$-effect or turbulent transport:} the
transport of the weak poloidal field from its generating region to the
region of strong shear, the tachocline. ~~{$(c)$} {The
$\omega$-effect:} the organization and amplification of the magnetic
field by differential rotation, particularly by large-scale rotational
shear in the tachocline, into strong, isolated magnetic structures that
are toroidal in character.  ~~{$(d$)} {Magnetic buoyancy:}  the rise
and transport of the large-scale toroidal field by magnetic buoyancy
into and through the convection zone to be either shredded and recycled
or to emerge as active regions.

Since all models presume close linkages between the differential
rotation of the sun and the operation of its global dynamo, let us
briefly review what is known about the angular velocity $\Omega$
profile with radius and latitude.  Helioseismology, which involves the
study of the acoustic $p$-mode oscillations of the solar interior
(e.g. Gough \& Toomre 1991), has provided a new window for studying
dynamical processes deep within the sun.  This has been enabled by the
nearly continuous and complementary helioseismic observations provided
from both the vantage point of the SOHO spacecraft with the
high-resolution Michelson Doppler Imager (SOI--MDI) (Scherrer et al.
1995) and from the ground-based Global Oscillation Network Group
(GONG) set of six related instruments distributed at different
longitudes across the Earth (Harvey et al. 1996).
Helioseismology has revealed that the rotation profiles obtained by
inversion of frequency splittings of the $p$ modes (e.g., Thompson et
al. 1996, Schou et al. 1998, Howe et al. 2000, Thompson et al. 2003)
have a striking behavior that is unlike any anticipated by convection
theory prior to such probing of the interior of a star.  The strong
latitudinal variation of angular velocity $\Omega$ observed near the
surface, where the rotation is considerably faster at the equator than
near the poles, extends through much of the convection zone depth
(about 200 Mm) with relatively little radial dependence.
Another striking feature is the tachocline (e.g., Spiegel \& Zahn
1992), a region of strong shear at the base of the convection zone
where $\Omega$ adjusts to apparent solid body rotation in the deeper
radiative interior.  A thin {\sl near-surface shear layer} is also
present in which $\Omega$ increases with depth at intermediate and low
latitudes.  This subsurface region is now being intensively probed
using local domain helioseismic methods, revealing the presence of
remarkable large-scale meandering flow fields much like jet streams,
banded zonal flows and evolving meridional circulations, all of which
contribute to what is called Solar Subsurface Weather (SSW) (Haber et
al. 2000, 2002; Toomre 2002).

\subsection{Studying Elements of the Global Dynamo}

Computational resources are currently insufficient to enable modelling
a complete dynamical system incorporating all the diverse aspects of
the large-scale solar dynamo.  Our goal is therefore to study
individually some of the essential processes, with a view to eventually
combine such findings into a more complete nonlinear interface-type
solar dynamo model as resources become available.  In this spirit,
there has been substantial theoretical progress recently in trying to
understand how the differential rotation profiles deduced from
helioseismology may be established in the bulk of the convection zone.
Building on the early three-dimensional numerical simulations of
rotating convection in spherical shells (e.g., Gilman \& Miller 1981;
Glatzmaier \& Gilman 1982; Glatzmaier 1985a,b, 1987; Sun \& Schubert 1995),
recent modelling using the anelastic spherical harmonic (ASH) code on
massively-parallel supercomputers (e.g., Miesch et al.  2000; Elliott,
Miesch \& Toomre 2000; Brun \& Toomre 2002) has permitted
attaining fairly turbulent states of convection in which the resulting
$\Omega$ profiles now begin to capture many elements of the deduced
interior profiles.  These simulations possess fast equatorial rotation,
substantial contrasts in $\Omega$ with latitude, and reduced tendencies
for $\Omega$ to be constant on cylinders.  The role of the Reynolds
stresses and of meridional circulations within such convection in
redistributing the angular momentum to achieve such differential
rotation over much of the convection zone is becoming evident. However,
the simulations with ASH have only just begun to examine how the
near-surface rotational shear layer may be established (DeRosa, Gilman
\& Toomre 2002), whereas the formation and maintenance of a tachocline
near the base of the convection zone has only been tentatively
considered within three-dimensional simulations that admit downward
penetration (Miesch et al. 2000). 

Dynamics within the solar tachocline and overshoot region are thought
to be extremely complex (e.g.\ Gilman 2000; Ossendrijver 2003).  The
upper portion of the tachocline may extend into the convective
envelope whereas the lower portion consists of a stably-stratified,
magnetized shear flow.  Turbulent penetrative convection transfers
mass, momentum, energy, and magnetic fields between the convection
zone and radiative interior both directly and through the generation
of internal waves, particularly gravity waves, which can drive
oscillatory zonal flows and large-scale circulations.  Instabilities
driven by shear and magnetic buoyancy further influence the structure
and evolution of the tachocline and likely play an important role in
the solar activity cycle.  Understanding these various processes will
require much future work beyond the scope of this paper.

Our objective here is to expand upon the purely hydrodynamical
simulations with ASH to begin to study the magnetic dynamo action that
can be achieved by global-scale turbulent flows within the bulk of the
solar convection zone.  These studies build on the pioneering
modelling that was able to resolve fairly laminar but intricate
magnetohydrodynamic (MHD) convection and its dynamo action within
rotating spherical shells (e.g., Gilman \& Miller 1981, Gilman 1983,
Glatzmaier 1987).  Other related dynamo simulations have also
considered deeper shells (e.g., Kageyama, Watanbe \& Sato 1995,
Kageyama \& Sato 1997).  We turn now to more complex states associated
with the turbulent flows that can be resolved using the ASH code.
Much as in Brun \& Toomre (2002) and its immediate progenitors, we
will deal primarily with the bulk of the convection zone by imposing
stress-free and impenetrable upper and lower boundaries to the shell,
thereby ignoring the region of penetration of flows into the deeper
radiative interior.  Thus issues concerning the tachocline are not
dealt with, including the downward transport of magnetic fields
($\beta$-effect) into this region where strong toroidal fields may be
stretched into existence.  Likewise the stability of these fields
and the buoyant rise and emergence of flux tubes is not studied in detail,
although magnetic buoyancy is allowed in our ASH simulations via the
anelastic approximation.  Rather, the simulations reported here 
examine the $\alpha$-effect and $\omega$-effects within
much of the convective interior, inspired particularly by the Gilman
\& Miller (1981) studies, but now having the ability to resolve
turbulent convection and the fairly realistic differential rotation
that it is able to sustain.

The convection in many previous studies of dynamo action in rotating
spherical shells is dominated by so-called banana cells: columnar rolls
aligned with the rotation axis.  These cells possess substantial
helicity and generally drive a large differential rotation,
thus providing all the necessary ingredients for an $\alpha-\omega$ 
dynamo.  Sustained dynamo action is indeed observed for a variety of
parameter regimes, but the results are generally not solar-like.
The first studies by Gilman \& Miller (1981) revealed no solutions 
with periodic field reversals.  Cyclic, dipolar dynamos were found by 
Gilman (1983) and Glatzmaier (1984, 1985a,b) for somewhat higher 
Rayleigh numbers but the periods were significantly shorter than the
solar activity cycle ($\sim$ 1--10 yrs) and toroidal fields were 
found to propagate poleward during the course of a cycle rather 
than equatorward as in the sun.  Furthermore, these relatively
low-resolution simulations could not capture the intricate structure 
of the fluctuating field components known to exist in the 
solar atmosphere.

More recent simulations of MHD convection in rotating spherical shells
have generally focused on parameter regimes more characteristic of the
geodynamo and other planetary interiors (e.g.\ Kageyama \& Sato 1997;
Christensen, Olson \& Glatzmaier 1999; Roberts \& Glatzmaier 2000;
Busse 2000a,b; Ishihara \& Kida 2002).  Relative to the sun,
convective motions in the planetary interiors are much more influenced
by rotation (lower Rossby numbers) and diffusion (lower Reynolds 
and magnetic Reynolds numbers) and much less influenced by
compressibility (mild density stratification).  Although such
simulations have achieved higher resolution relative to Gilman and
Glatzmaier's earlier work, they are still generally dominated by
banana cells due to the strong rotational influence.  They often tend
to produce mean fields of a dipolar nature, although quadrupolar
configurations are preferred in some parameter regimes, generally
characterized by high Rayleigh numbers and low magnetic Prandtl
numbers (Grote, Busse \& Tilgner 1999, 2000; Busse 2000b).  Cyclic
solutions have been found, but field reversals are more often
aperiodic, particularly for high Rayleigh numbers.

In this paper we report simulations of hydromagnetic dynamo action in
the solar convection zone at unprecedented spatial resolution.  Our
primary objective is to gain a better understanding of magnetic field
amplification and transport by turbulent convection in the solar
envelope and the essential role that such processes play in the
operation of the solar dynamo.  In the \S2 we
describe our numerical model and our simulation strategy in which we
introduce a small seed magnetic field into an existing hydrodynamic
simulation.  In \S3 we discuss some properties of this hydrodynamic
progenitor simulation and the exponential growth and nonlinear
saturation of the seed field.  We then investigate the intricate
structure and evolution of the dynamo-generated fields in \S4 and
their back-reaction on mean flows in \S5.  Here we shall focus
on the turbulent or fluctuating (non-axisymmetric) field components
which are found to dominate the magnetic energy.  We consider the mean
(axisymmetric) field components separately in \S6.  In \S7 we discuss
the magnetic and kinetic helicity found in our dynamo simulations and
present spectra and probability density functions for various fields.
We summarize our primary results and conclusions in \S8.

\section{Modelling Approach}

\subsection{Anelastic MHD Equations}\label{equations}

In this paper we report three-dimensional numerical experiments
designed to investigate the complex magnetohydrodynamics (MHD) of the
solar convection zone in spherical geometries.  We have extended our
already well-tested hydrodynamic ASH code (anelastic spherical
harmonic; see Clune et al. 1999, Miesch et al. 2000, Brun \& Toomre
2002) to include the magnetic induction equation and the feedback of
the field on the flow via Lorentz forces and ohmic heating. Thus, the
ASH code is now able to solve the full set of 3--D MHD anelastic
equations of motion in a rotating, convective spherical shell
(Glatzmaier 1984) with high resolution on massively-parallel computing
architectures.  These equations are fully nonlinear in velocity and
magnetic field variables, but under the anelastic approximation the
thermodynamic variables are linearized with respect to a spherically
symmetric and evolving mean state having a density $\rb$, pressure
$\bar{P}$, temperature $\tb$ and specific entropy $\sb$.  Fluctuations
about this mean state are denoted by $\rho$, $P$, $T$, and $S$.  The
resulting equations are:

\begin{eqnarray}
\nab\cdot(\rb {\bf v}) &=& 0, \\
\nab\cdot {\bf B} &=& 0, \\
\rb \left(\frac{\p {\bf v}}{\p t}+({\bf v}\cdot\nab){\bf v}+2{\bf \Omega_o}\times{\bf v}\right) 
 &=& -\nab P + \rho {\bf g} + \frac{1}{4\pi} (\nab\times{\bf B})\times{\bf B} \nonumber \\
&-& \nab\cdot\mbox{\boldmath $\cal D$}-[\nab\bar{P}-\rb{\bf g}], \\
\rb \tb \frac{\p S}{\p t}+\rb \tb{\bf v}\cdot\nab (\sb+S)&=&\nab\cdot[\kappa_r \rb c_p \nab (\tb+T)
+\kappa \rb \tb \nab (\sb+S)]\nonumber \\
&+&\frac{4\pi\eta}{c^2}{\bf j}^2+2\rb\nu\left[e_{ij}e_{ij}-1/3(\nab\cdot{\bf v})^2\right]
 + \rb {\epsilon},\\
\frac{\p {\bf B}}{\p t}&=&\nab\times({\bf v}\times{\bf B})-\nab\times(\eta\nab\times{\bf B}),
\end{eqnarray}
where ${\bf v}=(v_r,v_{\theta},v_{\phi})$ is the local velocity in spherical coordinates in 
the frame rotating at constant angular velocity ${\bf \Omega_o}$, ${\bf g}$ is the 
gravitational acceleration, ${\bf B}=(B_r,B_{\theta},B_{\phi})$ is the magnetic field, 
${\bf j}=c/4\pi\, (\nab\times{\bf B})$ is the current density, 
$c_p$ is the specific heat at constant pressure, $\kappa_r$ is the radiative diffusivity, $\eta$ is the 
effective magnetic diffusivity, and ${\bf \cal D}$ is the viscous stress tensor, involving the components
\begin{eqnarray}
{\cal D}_{ij}=-2\rb\nu[e_{ij}-1/3(\nab\cdot{\bf v})\delta_{ij}],
\end{eqnarray}
where $e_{ij}$ is the strain rate tensor, and $\nu$ and $\kappa$ are effective eddy diffusivities. 
A volume heating term $\rb \epsilon$ is also included in these equations for completeness but
it is insignificant in the solar envelope.  When our model is applied to other stars, 
such as A-type stars (Browning, Brun \& Toomre 2004), this term represents energy 
generation by nuclear burning.
To complete the set of equations, we use the linearized equation of state
\begin{equation}
\frac{\rho}{\rb}=\frac{P}{\bar{P}}-\frac{T}{\tb}=\frac{P}{\gamma\bar{P}}-\frac{S}{c_p},
\end{equation}
where $\gamma$ is the adiabatic exponent, and assume the ideal gas law 
\begin{eqnarray}
\bar{P}={\cal R} \rb \tb
\end{eqnarray}
where ${\cal R}$ is the gas constant.  The reference or mean state
(indicated by overbars) is derived from a one-dimensional solar
structure model (Brun et al.\ 2002) and is continuously updated with
the spherically-symmetric components of the thermodynamic fluctuations
as the simulation proceeds.  It begins in hydrostatic balance so the
bracketed term on the right-hand-side of equation (3) initially
vanishes.  However, as the simulation evolves, turbulent and magnetic
pressure drive the reference state slightly away from hydrostatic
balance.

Due to limitations in computing resources, no simulation achievable
now or in the near future can hope to directly capture all
scales of solar convection from global to molecular dissipation
scales.  The simulations reported here resolve nonlinear interactions
among a larger range of scales than any previous MHD model of
global-scale solar convection but motions still must exist in the sun
on scales smaller than our grid resolution.  In this sense, our
models should be regarded as large-eddy simulations (LES) with
parameterizations to account for subgrid-scale (SGS) motions.  Thus
the effective eddy diffusivities $\nu$, $\kappa$, and $\eta$ represent
momentum, heat, and magnetic field transport by motions which are not
resolved by the simulation.  They are allowed to vary with radius but
are independent of latitude, longitude, and time for a given 
simulation.  Their amplitudes and radial profiles are varied depending
on the resolution and objectives of each simulation.  In the simulations
reported here, $\nu$, $\kappa$, and $\eta$ are assumed to be proportional
to $\rb^{-1/2}$.

The velocity, magnetic, and thermodynamic variables are expanded in
spherical harmonics $Y_{\ell m}(\theta,\phi)$ for their horizontal
structure and in Chebyshev polynomials $T_n(r)$ for their radial
structure (see Appendix A).  This approach has the advantage that the
spatial resolution is uniform everywhere on a sphere when a complete
set of spherical harmonics is used up to some maximum in degree $\ell$
(retaining all azimuthal orders $m \leq \ell$ in what is known as
triangular truncation).

The anelastic approximation captures the effects of density
stratification without having to resolve sound waves which would
severely limit the time step.  In the MHD context, the anelastic
approximation filters out fast magneto-acoustic waves
but retains the Alfven and slow magneto-acoustic
modes.  In order to ensure that the mass flux and the magnetic field
remain divergenceless to machine precision throughout the simulation,
we use a toroidal--poloidal decomposition as:
\begin{eqnarray}
{\rb\bf v}=\nab\times\nab\times (W \hat{\bf e}_r) +  \nab\times (Z \hat{\bf e}_r), \\ 
{\bf B}=\nab\times\nab\times (C \hat{\bf e}_r) +  \nab\times (A \hat{\bf e}_r) ~~~. 
\end{eqnarray}

Appendix A lists the full set of anelastic MHD equations as solved by
the numerical algorithm, involving the spherical harmonic coefficients 
of the streamfunctions $W$ and $Z$ and the magnetic potentials $C$ 
and $A$.  This system of equations requires 12 boundary conditions in
order to be well-posed. Since assessing the angular momentum
redistribution in our simulations is one of the main goals of this
work, we have opted for torque-free velocity and magnetic boundary
conditions:
\begin{enumerate}
\item impenetrable top and bottom: $v_r=0|_{r=r_{bot},r_{top}}$,
\item stress free top and bottom: $\frac{\p}{\p r}\left(\frac{v_{\theta}}{r}\right)=\frac{\p}{\p r}\left(\frac{v_{\phi}}{r}\right)=0|_{r=r_{bot},r_{top}}$,
\item constant entropy gradient at top and bottom: $\frac{\p \sb}{\p r}=cst|_{r=r_{bot},r_{top}}$
\item match to an external potential magnetic field at top and bottom: $\BB=\nab \Phi => \Delta \Phi =0|_{r=r_{bot},r_{top}}$,
or impose a purely radial magnetic field at top and bottom (match to a highly permeable external media, Jackson 1999), i.e $B_{\theta}=B_{\phi}=0|_{r=r_{bot},r_{top}}$  
\end{enumerate}
The main difference between having a potential or a purely radial
magnetic field is that with the latter the Poynting flux is zero at the
shell surface, and thus there is no leakage of magnetic energy through the
boundaries (see \S3.2).


\subsection{Numerical Experiments}

Our numerical model is a simplified portrayal of the solar convection
zone: solar values are taken for the heat flux, rotation rate, mass
and radius, and a perfect gas is assumed.  The computational domain
extends from 0.72 to 0.97 $R_*$ (with $R_*$ the solar radius), thereby
focusing on the bulk of the unstable zone without yet considering
penetration into the radiative interior or smaller-scale convective
motions near the photosphere.  The depth of the convection zone is
therefore $L=1.72\times 10^{10}$ cm and the background density varies
across the shell by about a factor of 30.  Outward heat transport by
unresolved convective motions near the surface is modeled by locally
increasing the component of the subgrid-scale (SGS) eddy diffusivity
$\kappa$ which operates on the mean (horizontally-averaged) entropy
gradient, thus allowing the simulation to achieve flux equilibrium
(see \S3.2).  Meanwhile, the influence of unresolved motions on the
flow itself is taken into account through the SGS eddy diffusivities
$\nu$, $\kappa$, and $\eta$.

The magnetic simulations discussed here were all initiatiated
from the same non-magnetic progenitor simulation which we refer to as 
case H.  Case H is well-evolved, with a complex convective
structure and a solar-like differential rotation profile (\S3.1).

A small seed magnetic field is then introduced and its evolution
is followed via the induction equation.  The seed field is dipole 
in nature but soon develops a more complicated structure as it is amplified 
by the convective motions.  If the magnetic diffusivity is sufficiently
small, the field will continue to amplify until it reaches a nonlinear
saturation level where production balances dissipation.   In order to
determine whether sustained dynamo action is achieved, the simulation
must be evolved for at least several ohmic diffusion times 
$\tau_{\eta}=L^2/(\pi^2 \eta)$
(see Moffatt 1978, Jacobs 1987). 
We have conducted three MHD simulations, cases {\em M1}, {\em M2} and {\em M3} each with 
progressively lower values of the magnetic diffusivity (see Table 1).

It is currently impractical to perform dynamo calculations with a
spatial resolution comparable to our most turbulent hydrodynamic cases
($N_\theta=1024$, $N_\phi=2048$, $N_r=256$) which achieve an rms
Reynolds number $R_e$ of over 700.  The increased workload required to
solve the magnetic induction equation and the long time integrations
necessary to reliably assess dynamo action cannot be easily achieved
with currently available computational resources.  In order to achieve
dynamo action in more moderately turbulent simulations such as those
considered here ($R_e \sim 150$), the magnetic Prandtl number
$P_m=\nu/\eta$ must be greater than unity, whereas in the sun it is
significantly less than unity (based on microscopic values for $\nu$ and
$\eta$).  This is a well known difficulty in dynamo simulations within 
astrophysical or geophysical contexts (see for example Christensen et
al. 1999).  However, the diffusivities in our simulations arise from
unresolved convective motions, not microscopic processes, and the
effective transport properties of such motions are thought to yield
Prandtl and magnetic Prandtl numbers of order unity.


\section{Convection, Rotation, and the Generation of Fields}

\subsection{Progenitor Non-Magnetic Convection}

Figure 1 illustrates the convective structure and differential
rotation for the hydrodynamical progenitor case H immediately prior
to introducing a seed magnetic field.  The radial velocity near the
top of the domain is shown using a Mollweide projection which displays
the entire horizontal layer with minimal distortion.  The circular
arcs ($\pm 90^{\circ}$) encompass a hemisphere and the rest of the
globe is contained in the lunes on either side.  The convection
patterns are complex, time-dependent, and asymmetric due to the
density stratification, consisting of relatively weak, broad upflows
with narrow, fast downflows around their periphery.  This asymmetry
translates into a net downward transport of kinetic energy. The strong
correlations between warm upward motions and cool downward motions are
essential in transporting heat outward.


There is a clear difference in the size and structure of the
convective patterns at low and high latitudes.  Near the equator the
downflow lanes tend to align with the rotation axis in the north/south
direction whereas at higher latitudes ($\gtrsim 25^{\circ}$) they tend
to be more isotropic and of smaller spatial extent.  Part of this
behavior can be understood by considering the cylinder which is
aligned with the rotation axis and tangent to the inner boundary.
This tangent cylinder intersects the outer boundary at latitudes of
about $42^{\circ}$. It is well known that in a rotating convective
shell the flow dynamics are different inside and outside of the inner
tangent cylinder (Busse 1970, Busse \& Cuong 1977).  The connectivity
of the flow, the influence of Coriolis forces, and the distance to the
rotation axis are different in the polar regions relative to the
equatorial regions, leading to different convective patterns in midly
turbulent simulations such as case H. At low Reynolds numbers the transition
between equatorial modes and polar modes occurs near the tangent cylinder.
As the Reynolds number is increased this transition moves to lower 
latitudes and becomes less apparent.  For example, Brun \& Toomre (2002) have
demonstrated that increasing the level of turbulence in the simulations makes
the convective patterns in the equatorial region more isotropic and
extended downflow lanes become difficult to isolate within the
convective network.

\begin{figure}[!ht]
\setlength{\unitlength}{1.0cm}
\begin{picture}(5,5)
\includegraphics{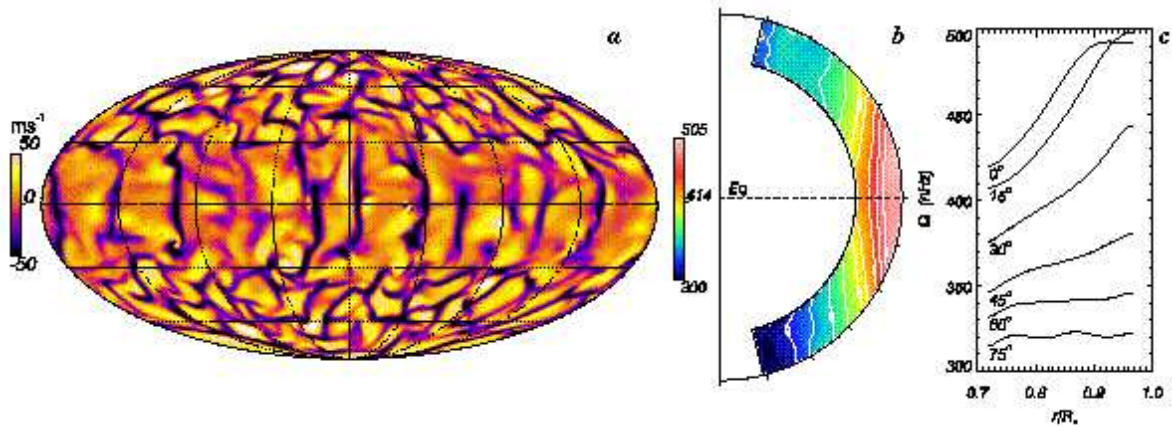}
\end{picture}
\caption[]{\label{fig1} The radial velocity near the top of the shell
for case H is shown in ($a$) using a Mollweide projection.  Dashed
lines indicate the equator as well as meridians and parallels every
45$^\circ$ and 30$^\circ$ respectively.  Downflows appear dark and
upflows bright.  Frame ($b$) illustrates the angular velocity $\Omega$
in case H averaged over longitude and time, with brighter tones
indicating more rapid rotation (see color tables).  In
frame ($c$) the mean angular velocity is shown as a function of radius
for the indicated latitudes, averaged over both hemispheres.}
\end{figure}

Vortical plumes are evident at the interstices of the downflow
network, representing coherent structures which are surrounded by more
chaotic flows.  The sense of the vorticity is generally cyclonic;
counterclockwise in the northern hemisphere and clockwise in the
southern hemisphere. The strongest downflow plumes
extend through the entire depth of the domain.  They tend to align
with the rotation axis and to tilt away from the meridional plane,
leading to Reynolds stresses that are crucial ingredients in
redistributing the angular momentum within the shell (cf. \S 5, see
also Miesch et al. 2000, Brun \& Toomre 2002).  Downflow lanes and
plumes are continually advected, sheared, and distorted by
differential rotation and nonlinear interactions with other flow
structures.

The differential rotation in case H is shown in Figures 1$b$ and 1$c$,
expressed in terms of the sidereal angular velocity $\Omega$.  The
angular velocity of the rotating reference frame is 414 nHz, which
corresponds to a rotation period of 28 days.  In the contour plot, the
polar regions have been omitted due to the difficulty of forming
stable averages there as a result of the small moment arm and small
averaging domain.

Case H exhibits a differential rotation profile which is in good
agreement with the solar internal rotation profile inferred from
helioseismology in the bulk of the convection zone (Thompson et al.\
2003).  Angular velocity contours at mid-latitudes are nearly radial
and the rotation rate decreases monotonically with increasing latitude
as in the sun.  The latter property in particular represents an
important improvement over most previous spherical convection
simulations in which the latitudinal angular velocity contrast
$\Delta\Omega$ was confined mainly to low and mid-latitudes, namely
outside of the inner tangent cylinder.  The angular velocity profile
in such simulations is generally sensitive to the parameters of the
problem, and more solar-like profiles such as case H can be achieved
by varying the Reynolds and Prandtl numbers in particular (Elliott,
Miesch \& Toomre 2000; Brun \& Toomre 2002).  The differential
rotation contrast between the equator and latitudes of $60^\circ$ in case H is
140 nHz (or 34\% relative to the frame of reference), somewhat larger
than the 92 nHz (or 22\%) variation implied by helioseismology. The
rotation profile of case H exhibits some asymmetry with respect to the
equator, particularly at high latitudes (Fig.\ 1$b$), although such
asymmetries are expected to diminish over a longer temporal average.
Since the convection itself is generally asymmetric, it is not
surprising that the mean flows driven by the convection are as well.

Mean-field models of the solar differential rotation have advocated
that a thermal wind balance (involving latitudinal temperature
gradients) may be the cause of the non-cylindrical angular velocity
profile (Kichatinov \& R\"udiger 1995, Durney 1999).  This may come
about if baroclinic convective motions produce latitudinal heat flux,
leading to a breakdown of the Taylor-Proudman theorem (Pedlosky 1987).
A pole-equator temperature contrast of few degrees K is compatible
with a $\Delta\Omega/\Omega_o$ of $\sim 30\%$.  Although it is indeed
true that case H exhibits latitudinal entropy and temperature
gradients, these are not the dominant players in driving the differential
rotation throughout the shell.
Rather, we find that the Reynolds stresses are the main agents
responsible for maintaining the rotation profiles in our simulations
(see \S5).

\subsection{Achieving Sustained Dynamo Action}

We now consider the dynamo possibilities that such intricate
convective patterns and large differential rotation can lead to.  As
stated earlier, we have introduced a seed magnetic poloidal field into
our hydrodynamical case H for three different values of the magnetic
diffusivity $\eta$, corresponding to cases {\em M1}, {\em M2} and {\em M3} (Table
1).  Figure 2 shows the magnetic and kinetic energy evolution for
these three cases.  We note that over more than 4000 days
(corresponding to several ohmic decay times, cf. Table 1) the two
least diffusive cases {\em M2} and {\em M3} achieve a sustained magnetic
energy (ME), the amplitude of which depends on $\eta$.  The initial
exponential growth of ME in case {\em M3} lasts for about 600 days, after
which the nonlinear feedback of the Lorentz forces on the flow begins
to saturate the dynamo. For case {\em M2} which has a slower growth rate,
another linear phase seems to last for at least 4000 days and it is
unclear whether it has truly saturated.  By contrast, case {\em M1} is
clearly decaying, since the rate of generation of magnetic fields in
the entire shell volume ($\int_V {\bf v}\cdot[{\bf B}\times{\bf j}] dV$)
cannot compensate for the rate of destruction by ohmic diffusion
($\int_V 4\pi\eta/c \, {\bf j}^2 dV$).  Interpolating between cases {\em M1}
and {\em M2} to find the zero growth rate yields a critical magnetic
diffusivity at mid depth $\eta\sim 5.9\times 10^{11}$ cm s$^{-2}$.  In
terms of the magnetic Reynolds number (see Table 1), we find that
$R_m$ must be at least $300$ for sustained dynamo action to
occur.  This value of $R_m$ is about 25\% larger than in the
incompressible simulations of Gilman (1983) which consider 
a simpler configuration.

\begin{figure}[!ht]
\setlength{\unitlength}{1.0cm}
\begin{picture}(5,7)
\includegraphics{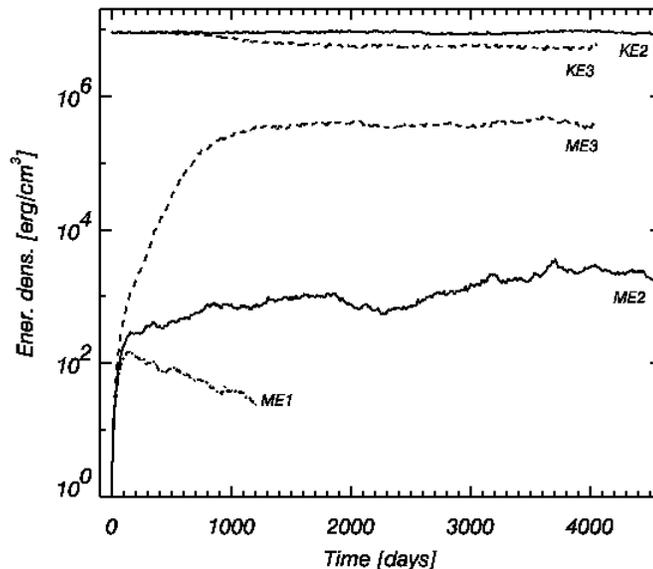}
\end{picture}
\caption[]{\label{fig2} The temporal evolution of the volume-integrated kinetic 
energy density (KE) and magnetic energy density (ME) 
is shown for cases {\em M1} ($R_m=272$), {\em M2} ($R_m=334$) and {\em M3} 
($R_m=486$), represented by dot-dashed, solid, and dashed lines respectively.}
\end{figure}

Upon saturation, the kinetic energy (KE) in model {\em M3} has been
reduced by about 40\% compared to its initial value, say KE$_0$, given by
case H (see Table 2).  This change is mostly due to a reduction of
the energy contained in the differential rotation (DRKE) which drops
by over 50\%. By contrast, the energy contained in the convective
motions (CKE) only decreases by about 27\%, which implies an increased
contribution of the non-axisymmetric motions to the total kinetic
energy balance. For case {\em M3}, the decrease in KE first becomes
apparent after about 600 days of evolution, when the ME reaches
roughly 0.5\% of KE$_0$.  After 1200 days, the ME reaches a value of
about 8\% of the KE and retains that level for more than 3 ohmic decay
times $\tau_{\eta}$.  The ME in case {\em M2} is still too small ($\leq
0.1\%$) even after 4000 days for Lorentz forces to have a significant
influence on the convective motions, as demonstrated by comparing the
kinetic energy evolution in cases {\em M2} and {\em M3}.

It is instructive to briefly consider the exchange of energy among
different reservoirs in our simulations.  We refer to Starr \& Gilman
(1966) for a more detailed discussion of energy exchange in an MHD
system. We first note that both the total kinetic and magnetic
energies remain small compared to the total potential, internal and
rotational energies contained in the shell.  Further, the magnetic
energy must arise from the conversion of kinetic energy but this does
not necessarily lead to a decrease in the total kinetic energy because
the motions may draw upon other reservoirs.  Yet, in all of our
magnetic simulations, energy is redistributed such that the sum of the
kinetic and magnetic energy is less than the total kinetic energy
contained in case H.  The net energy deficit can be attributed
primarily to the reduction in strength of the differential rotation by
Maxwell stresses.  This means that in a convection zone the way the
energy is redistributed among and within the different reservoirs is
modified by the presence of magnetic field, but these modifications
remain small in the cases presented here. We refer to Cattaneo, Emonet
\& Weiss (2002) for a detailed study of the influence of an imposed 
magnetic field on Boussinesq convection.

\begin{figure}[!ht]
\setlength{\unitlength}{1.0cm}
\begin{picture}(5,5.4)
\includegraphics{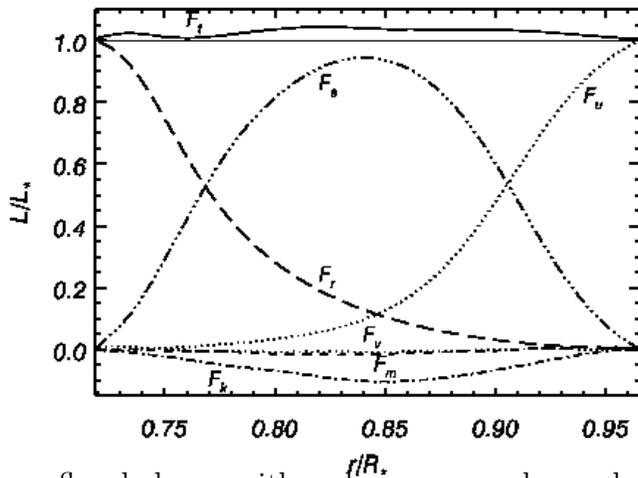}
\end{picture}
\caption[]{\label{fig3} Energy flux balance with radius, averaged over 
horizontal surfaces and in time. The net radial energy flux in case {\em M3}
(solid line) is expressed as an integrated luminosity
through horizontal shells and normalized with respect to the solar
luminosity, $L_*$.  In the other curves, this net flux is
separated into components as defined in equations (11)--(17),
including the enthalpy flux $F_e$, the radiative flux $F_{\rm r}$, 
the unresolved eddy flux $F_u$, the kinetic energy flux  $F_k$,
the Poynting flux $F_m$, and the viscous flux $F_v$.
}
\end{figure}

To further investigate the role played by the different agents in
transporting energy, we illustrate in Figure 3 the contribution of 
various physical processes to the total radial energy flux through 
the shell, converted to luminosity and normalized to the solar 
luminosity.   The net luminosity, $L(r)$, and its components 
are defined as:

\begin{equation}
F_e+F_k+F_{\rm r}+F_u+F_v+F_m=\frac{L(r)}{4\pi r^2} ~~~,
\end{equation}
with
\begin{eqnarray}
F_e&=&\rb\, c_p\, \overline{v_r T'} \,,\\
F_k&=&\frac{1}{2}\, \rb\, \overline{v^2 v_r} \,,\\
F_r&=& -\kappa_{r}\, \rb\, c_p\, \frac{d\tb}{d r} \,, \\
F_u&=& -\kappa\, \rb\, \tb\, \frac{d\sb}{d r} \,, \\
F_v&=& -\overline{{\bf v}\cdot {\bf \cal D}} \,, \\
F_m&=& \frac{c}{4\pi}\, \overline{E_{\theta}B_{\phi}-E_{\phi}B_{\theta}} \,,
\end{eqnarray}
where ${\bf E}=4\pi\eta {\bf j} c^{-2} - ({\bf v}\times\BB) c^{-1}$ is the
electric current, $F_e$ the enthapy flux, $F_k$ the kinetic energy
flux, $F_{\rm r}$ the radiative flux, $F_u$ the unresolved eddy flux,
$F_v$ the viscous flux and $F_m$ the Poynting flux.  The unresolved
eddy flux $F_u$ is the heat flux due to subgrid-scale motions which,
in our LES-SGS approach takes the form of a thermal diffusion
operating on the mean entropy gradient.  Its main purpose is to
transport energy outward through the impenetrable upper boundary where
the convective fluxes $F_e$ and $F_k$ vanish and the remaining fluxes
are small.  It should not be mistaken with $F_{\rm r}$, which is the
flux due to radiative diffusion and which operates on the mean
temperature gradient.  The radiative diffusivity, $\kappa_r$ is
derived from a one-dimensional solar structure model (Brun et al 2002),
whereas the eddy diffusivity
$\kappa$ is chosen to model the effects of small-scale motions and to
ensure that the flow is well resolved.  There is an additional energy flux,
$F_v$, which arises from the subgrid-scale eddy viscosity, $\nu$.

If the simulation were in a thermally-relaxed state, the total flux
through each horizontal surface would be constant and equal to the
solar luminosity which is applied at the upper and lower boundaries:
$L(r) = L_*$.  Figure 3 indicates that the normalized net flux $L/L_*$
(solid line) is indeed close to unity, implying that the simulation is 
close to thermal equilibrium.

The enthalpy flux here carries up to 90\% of the solar
luminosity in the bulk of the convective zone and $F_{\rm r}$ and
$F_u$ carry the energy at respectively the bottom and top of the
domain where $F_e$ vanishes. The remaining fluxes $F_k$, $F_v$ and
$F_m$ are relatively small and negative in most of the domain. The
downward direction of the kinetic energy flux is due to the asymmetry
between the fast downflow lanes and the slower broad upflows. This
downward flux carries about 10\% of the solar luminosity and possess a
bigger amplitude than either $F_v$ or $F_m$. The low amplitude of $F_v$
confirms that in our simulations inertia dominates over viscous effects,
i.e.\ the Reynolds number in all cases is much greater than unity.
Similarly, the low amplitude of the Poynting flux confirms that
magnetic processes in case {\em M3} do play a role in the overall energy
transport but not to the point of significantly modifying the
flux balance established in the non-magnetic progentitor case H.  The
volume integrated $ME$ is about 10\% of $KE$; it would likely require
a much higher level of magnetism in order for the Poynting flux to
have a substantial influence on the net energy transport.

The Poynting flux $F_m$ is also influenced by our choice of magnetic
boundary conditions. In all the magnetic cases presented here we match
the computed field to an internal and external potential field at
every time step.  This leads to a non-zero electromagnetic flux
through the boundaries.  We have investigated the impact of such
magnetic energy ``leakage'' on the dynamo action by computing one case
in which the magnetic field was required to be purely radial at the
boundaries, yielding no net Poynting flux through the shell ($F_m=0$
at the top and bottom boundaries). The effect of closed as opposed to
open boundary conditions seems to be that in the former the magnetic
energy amplification is more efficient, with potentially a lower
dynamo threshold. But since in the solar case such magnetic energy
``leakage'' exists both at the bottom via for example turbulent
pumping (Tobias et al. 2001) and at the photosphere via for example
magnetic eruptions, we consider that our choice of boundary conditions
is reasonable for the solar dynamo problem.  We further believe that
open magnetic boundary conditions play a central role in regulating
the magnetic dynamo action in the convection zone, by providing an
outlet for the magnetic energy and also most likely for the magnetic
helicity.

\section{Convective and Magnetic Structures}

\subsection{Flow Patterns and Their Evolution}

The structure of the convection in simulation {\em M3} is illustrated in Figure 4.
The convective patterns are qualitatively similar to the hydrodynamic case
H, which can be seen by comparing the radial velocity field in the
upper left frame of Figure 4 to that shown in Figure 1.  Cases {\em M1} and {\em M2}
also exhibit similiar patterns because the magnetic fields in these
simulations never grow strong enough to exert a substantial influence 
on the global flow structure.  However, Lorentz forces in localized 
regions of case {\em M3} do have a noticable dynamical effect,  particularly
with regard to the evolution of strong downflow lanes where magnetic 
tension forces can inhibit vorticity generation.

\begin{figure}[!t]
\setlength{\unitlength}{1.0cm}
\begin{picture}(5,9)
\includegraphics{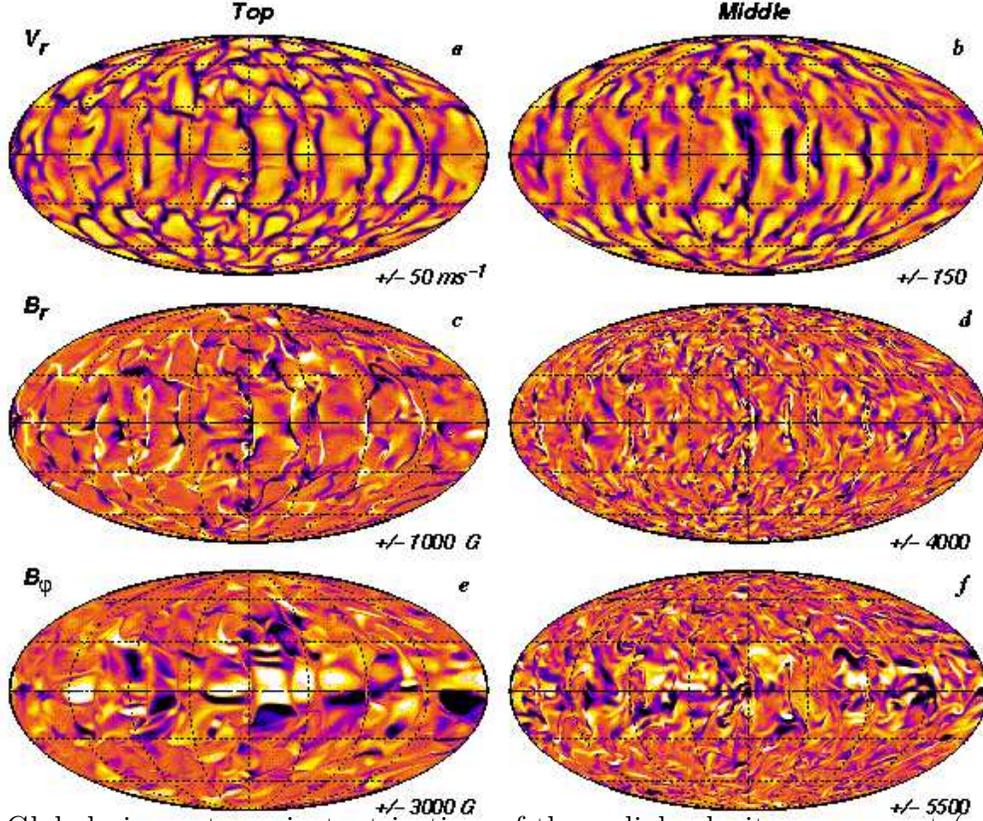}
\end{picture}
\caption[]{\label{slices} Global views at one instant in time of the radial 
velocity component (upper row) and the radial and longitudinal magnetic field 
components (middle and bottom rows) in case {\em M3} near the top (left column) and 
middle (right column) of the computational domain. Dark tones in turn
represent downflow, inward, and westward fields, with the ranges for each
color table indicated.  The color table is as in Figure 1.}
\end{figure}

The horizontal structure of the radial and longitudinal magnetic field
is also shown in Figure 4.  Many of the main features
are qualitatively similar to simulations of turbulent, compressible
magnetoconvection in Cartesian geometries (Cattaneo 1999; Stein \&
Nordlund 2000; Tobias et al.\ 2001).  The magnetic field generally has
a finer and more intricate structure than the velocity field due to
the smaller diffusion ($P_m=\nu/\eta=4$ in this simulation) and
also due to the nature of the advection terms in the induction
equation, which are similar in form to those in the vorticity equation
(e.g.\ Biskamp 1993).  Near the top of the shell, the radial magnetic
field $B_r$ is mainly concentrated in the downflow lanes, where both
polarities coexist in close proximity. By contrast, the toroidal field
$B_\phi$ near the surface appears more distributed and more patchy,
characterized by relatively broad regions of uniform polarity,
particularly near the equator.  The magnetic field topology generally
does not exhibit any clear symmetries about the equator, although some
of the $B_\phi$ patches at low latitudes do have an antisymmetric
counterpart.

In the middle of the shell the magnetic fluctuations appear of
smaller-scale and more distributed but they are still very
intermittent.  Strong vertical fields of mixed polarity still
correlate well with downflow lanes and plumes.  The longitudinal field
is more filamentary and is organized in longitudinally-elongated
structures, having been stretched by the gradients in angular velocity
(see also \S4.2).

\begin{figure}[!ht]
\setlength{\unitlength}{1.0cm}
\begin{picture}(5,5)
\includegraphics{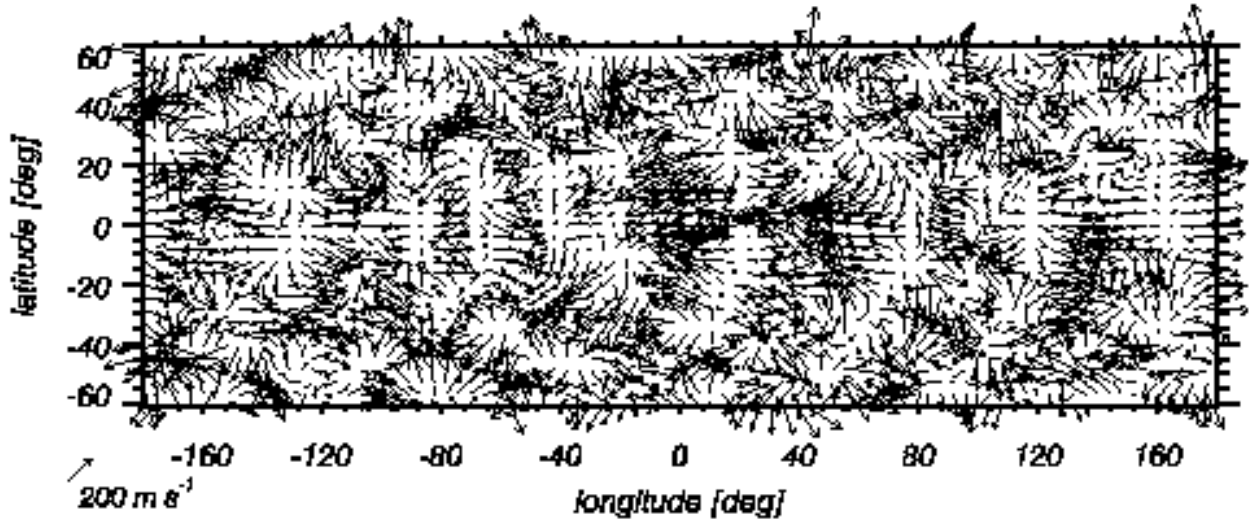}
\end{picture}
\caption[]{\label{fig5} Horizontal velocity vectors near the top of the
domain in case {\em M3} are shown for the latitude range $\pm 60\degrees$
at the same level and time as in Figure 4 (left column).
The axisymmetric velocity component has been subtracted out and the 
display grid is undersampled relative to the horizontal resolution 
of the simulation in order to improve clarity.  This sampling
does not capture more localized features such as vertically-aligned
vortex tubes which can be seen on higher-resolution images but which 
occur on scales smaller than the sampling grid.} 
\end{figure}

Throughout the shell, the magnetic field patterns evolve rapidly, as
fields are continuously transported, distorted, and amplified by
convective motions.  Of particular importance near the top of the
shell are the horizontal flows, shown in Figure 5 
for case {\em M3}.  Regions of convergence and divergence are apparent, as
are swirling vortices which occur most frequently at high and
mid-latitudes and generally have a cyclonic sense (counter-clockwise
in the northern hemisphere and clockwise in the southern).  Such flows
stretch the horizontal field and sweep the vertical field into
vortical downflow lanes where it is twisted, thus generating magnetic
helicity.  Horizontally-converging flows also squeeze together fields
of mixed polarity, driving magnetic reconnection.

The convective patterns visible in the vertical velocity field of
Figure 4 are also evident in the horizontal velocity patterns of Figure
5, particularly the dichotomy between low-latitudes which are
dominated by extended downflow lanes oriented north-south (visible
here as lines of horizontal convergence) and higher latitudes which
possess a smaller-scale, more isotropic downflow network.  If they
exist in the sun, such large-scale convective patterns may ultimately
be detectable in similar horizontal flow maps inferred from local-domain
helioseismic analyses using time-distance and ring-diagram procedures
(e.g.\ Haber et al.\ 2002; Hindman et al.\ 2003).  However, currently such
helioseismic flow maps are limited to the upper few percent of
the solar envelope, monitoring what is called solar sub-surface weather
(SSW; Toomre 2002).  This lies outside the computational domain
considered here.

\begin{figure}[!ht]
\setlength{\unitlength}{1.0cm}
\begin{picture}(5,15)
\includegraphics{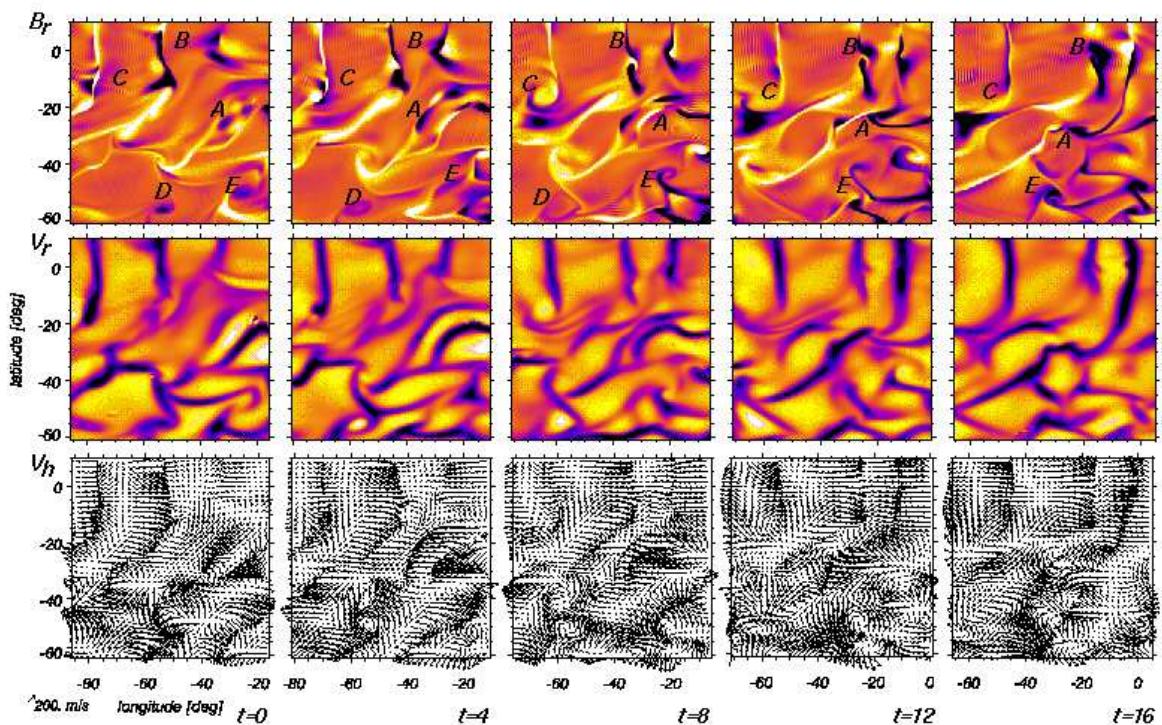}
\end{picture}
\caption[]{\label{fig6} The radial magnetic field ($B_r$, upper row), radial 
velocity ($v_r$, middle row), and horizontal flow vectors (${\bf v}_h$,
bottom row) are
shown for a selected horizontal domain near the top of the shell in
case {\em M3}.  A sequence of five snapshots is shown, each separated by
an interval of four days, with time increasing from left to right.
This region spans 70$\degrees$ in both latitude and longitude; 
the longitude range shifts eastward by $5\degrees$ with
each successive snapshot in order to track some of the flow features.
Particular features are indicated with labels.
The horizontal level and the time of the first
snapshot correspond to the global views shown in Fig.\ 4 (left column) and
Fig.\ 5.  The color table is the same as in Fig.\ 1.}
\end{figure}

The dynamical richness and rapid time evolution of the flow and
magnetic field patterns are highlighted in Figure 6.  The radial
magnetic field, the radial velocity field, and the horizontal flow all
exhibit an intricate structure which evolves substantially on time
scales of weeks and even days. Low-latitude features tend to drift
eastward relative to higher-latitude features due to advection by the
differential rotation and also due to inherent pattern propagation
relative to the local rotation rate.  At the tracking rate used in
Figure 6, this leads generally to a rightward movement of patterns
near the equator and a leftward movement of patterns near the southern
edge of the region shown (latitude -60$^\circ$).  In between,
particularly at a latitude of about $-25\degrees$, the distinctive
patterns at low and high latitudes meet, giving rise to a particularly
complex dynamical evolution.  The north-south aligned downflow lanes
at low latitudes temporarily link to the high-latitude network as they
drift by, and features caught in this interaction region are rapidly
sheared and distorted, forming filaments and vortices which then mix
and merge with other structures.

Figure 6 highlights the evolution of several features in particular,
indicated by letters.  The first of these, A, is a multi-polar region
which appears to represent several flux tubes passing through the
horizontal plane being visualized.  After they form, these localized
features are rapidly sheared by convective motions, distorting and
separating into flux sheets which then merge with other features
and lose their identity over the course of about two weeks.  Feature B
begins as a flux sheet confined to a north-south oriented downflow
lane where the polarity of the field is radially inward.  By the
second frame, flux of the opposite polarity (radially outward,
indicated by white) is advected into the downflow lane where it is
then wrapped up by the cyclonic vorticity and rapidly dissipated as it
reconnects with the existing field.  Similar dynamics are also
occurring in feature C which illustrates the merging of two flux sheets
of opposite polarity in a downflow lane (particularly evident in the
rightmost frame).  The lower portion of the outward-polarity sheet
(white) extends into the interface region at latitude -25$^\circ$,
where extended low-latitude downflow lanes merge with the
high-latitude network.  The intense vorticity and shear in this region
twist and stretch the field, dramatically changing its appearance on
a time scale of several days.

The most intense downflow plumes often possess enough vorticity to
evacuate the core of the plume due to centrifugal forces, and buoyancy
forces acting on the resulting decrease in density lead to a flow
reversal, creating a new upflow region which then diverges
horizontally due to the density stratification (Brandenburg et al. 1996, 
Brummell et al. 1998, Miesch et al. 2000). Such dynamics are occurring 
in feature D of Figure 6, now in the presence of a magnetic field.  
By the second frame, a new upflow is created in this manner (middle 
panels) which rapidly expands horizontally and interacts with the 
surrounding flow. There is a vertical flux tube present in the original 
downflow but it is rapidly dispersed as the flow reverses, losing all 
coherence by the fourth frame.  Feature E is another example of how 
field can be wrapped up by the vorticity in downflow lanes, particularly 
at high and mid-latitudes where the rotation vector has a large vertical 
component.

\subsection{Morphology of Magnetic Fields}\label{morphology}

\begin{figure}[!ht]
\setlength{\unitlength}{1.0cm}
\begin{picture}(5,13.5)
\includegraphics{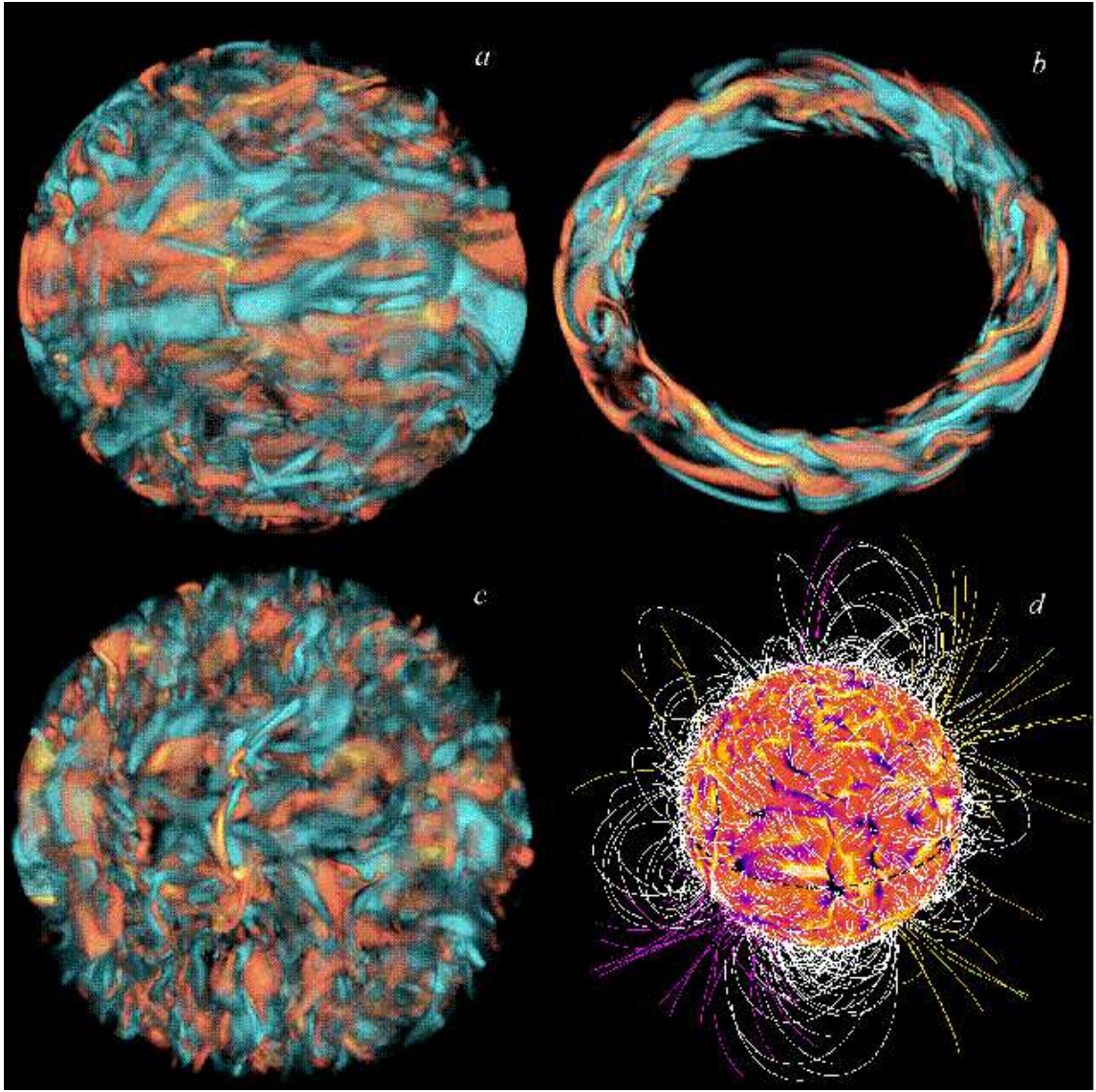} 
\end{picture}
\caption[]{\label{rendering} ($a$), ($c$) Volume renderings
of the toroidal ($B_\phi$) and radial ($B_r$) magnetic fields in case
{\em M3} at one instant in time (the same as in Fig.\ 4).  Red tones
indicate outward and eastward (prograde) fields and blue tones denote
inward and westward (retrograde) fields.  ($b$) shows a selected 
sub-volume of $B_\phi$ including the full span of longitude and radius
but only a narrow band in latitude centered around the equator.
The equatorial plane is tilted nearly perpendicular to the
viewing in order to highlight the radial and longitudinal structure.
Typical field strengths are about 1000 G for $B_r$ and 3000 G for $B_\phi$. 
($d$) Potential field extrapolation of the radial
magnetic field at the top of the computational domain.  The radial
field at the surface is shown in an orthographic projection and the
visualization traces individual field lines indicated in white
if they form closed loops and in yellow or magenta if they represent
open field of positive (outward) or negative (inward) polarity,
respectively. }
\end{figure}

Horizontal cross sections as in Figures 4--6 are informative but
they provide limited insight into the three-dimensional structure of
the flow and of the magnetic field in particular.  Further insight
requires volume visualizations as shown in Figure 7.

The toroidal and radial magnetic fields in Figures 7$a$ and 
7$c$ have a very different appearance, consistent with the 
contrast noted previously in Figure 4.  Whereas $B_r$ is concentrated into
vertically-oriented sheets and filaments, $B_\phi$ is organized into
relatively broad ribbons and tubes which extend mainly in longitude.
Figure 7$b$ further demonstrates the ribbon-like topology of the
toroidal field, showing in particular that the low-latitude horizontal
patches near the surface have a relatively small vertical extent,
although some meander in radius. Substantial magnetic helicity is
present throughout, involving complex winding of the toroidal field
structures along their length.  Some features resemble magnetic flux 
tubes but they generally do not remain coherent long enough for
magnetic buoyancy forces to induce them to rise.

Whereas some toroidal field structures maintain coherence over global
scales, the radial field is generally dominated by smaller-scale
fluctuations. In particular, radial field structures near the top of
the domain rarely penetrate deep into the convection zone, although
individual field lines maintain some connectivity throughout the
shell.  This connectivity also extends outside of the computational
domain because of the boundary conditions which match the interior
field to an external potential field.  The structure of this potential
field above the outer surface is illustrated in Figure 7{\it d}.  The
extrapolation shown in the figure treats the radial field near the top
of the domain as a source surface and requires that the field be
radial at 2.5 $R_*$, although field lines are only shown out 
to a radius of 1.5 $R_*$.

As in the sun, the surface magnetic field is complex, featuring
bipolar regions, nested loops, and an intricate web of connectivity
between both local and widely separated regions on the surface.
Although some large loops span both hemispheres, dipolar or
quadrupolar components are not evident and open field is not confined
to or even preferred in the polar regions.  Axisymmetric field
components are indeed present (see \S6), but the field morphology near
the surface and throughout the shell is dominated by smaller-scale
turbulent structures.

The magnetic energy in the potential field extrapolation decreases
rapidly with increasing radius, as spherical harmonic components decay
in proportion to $r^{-(l+1)}$.  A less dramatic outward gradient of
magnetic energy also occurs within the computational domain as
demonstrated in Figure 8.  Here we display the radial profile of the
total magnetic energy density integrated over the horizontal
dimensions after having broken it down into mean (axisymmetric) and
fluctuating (non-axisymmetric) poloidal and toroidal components in the
following manner:
\begin{eqnarray}
\mbox{MTE} &=&\frac{1}{8\pi}\left<B_{\phi}\right>^2 \,, \\
\mbox{MPE} &=&\frac{1}{8\pi}\left(\left<B_{r}\right>^2+\left<B_{\theta}\right>^2\right) \,, \\
\mbox{FTE} &=&\frac{1}{8\pi}\left((B_{\phi}-\left<B_{\phi}\right>)^2\right) \,, \\
\mbox{FPE} &=&\frac{1}{8\pi}\left((B_r-\left<B_{r}\right>)^2+(B_{\theta}-\left<B_{\theta}\right>)^2\right) \,, \\
\mbox{FME} &=&\frac{1}{8\pi}\left((B_r-\left<B_{r}\right>)^2+(B_{\theta}-\left<B_{\theta}\right>)^2+(B_{\phi}-\left<B_{\phi}\right>)^2\right) \,,
\end{eqnarray}
where the brackets $\left< ~ \right>$ denote a longitudinal
average. 

The magnetic energy generally peaks toward the bottom of the shell for
both the mean and fluctuating field components.  This is due in part
to the spherical divergence and the density stratification.  Downward
pumping of magnetic fields by convective motions also plays a role but
the pumping is not as effective as in penetrative convection
simulations where the underlying stable region provides a reservoir
where field can be accumulated and stored (cf.\ Tobias et al.\ 2001).

\begin{figure}[t]
\setlength{\unitlength}{1.0cm}
\begin{picture}(5,6)
\includegraphics{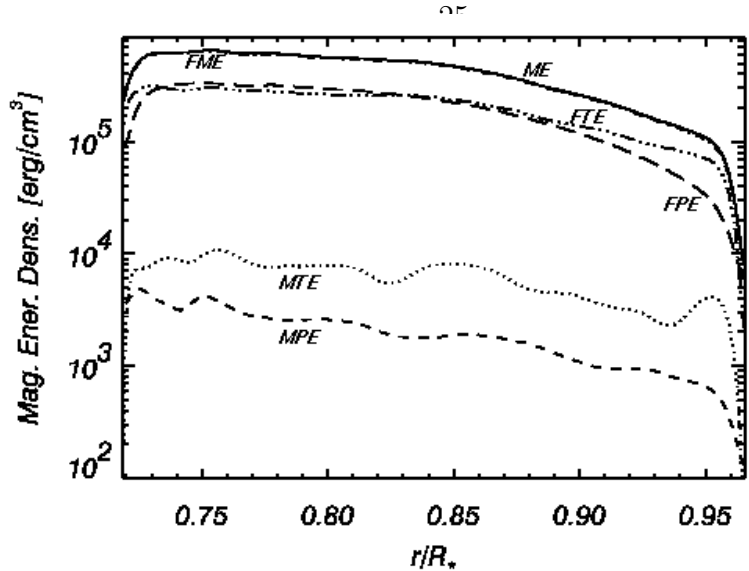}
\end{picture}
\caption[]{\label{meprofile} Radial profiles of the magnetic energy in case {\em M3}.
Shown are integrals over horizontal surfaces and averages in time of the
total magnetic energy (ME), the energy in the mean (axisymmetric) 
toroidal field (MTE) and the mean poloidal field (MPE), and the energy in the fluctuating
(non-axisymmetric) fields, including the toroidal component (FTE), the
poloidal component (FPE), and their sum (FME).}
\end{figure}

Figure 8 also shows that the magnetic energy contained in the
mean-field components is more than an order of magnitude smaller than
that contained in the non-axisymmetric fluctuations.  Most of the
mean-field energy is in the toroidal field, which exceeds the energy
in the poloidal field by about a factor of three due to the stretching
and amplification of toroidal field by differential rotation (the
$\omega$-effect).  This ratio is smaller than in the sun, where the
mean toroidal field is estimated to be about two orders of magnitude
more energetic than the mean poloidal field.  This discrepancy can
again be attributed to the absence of an overshoot region and
a tachocline, where toroidal field can be stored for extended periods
while it is amplified by relatively large angular velocity gradients
(see \S6).  For the non-axisymmetric fluctuations, the magnetic energy
is approximately equally distributed among the toroidal and poloidal
fields, indicating that the turbulent convection can efficiently
generate both components in roughly equal measure, implying that 
the $\omega$-effect plays a lesser role.

\section{Differential Rotation and Meridional Circulation}

Surface measurements and helioseismic inferences of large-scale,
axisymmetric, time-averaged flows in the sun currently provide the
most important observational constraints on global-scale models of
solar convection.  The structure, evolution, and maintenance of mean
flows (averaged over longitude and time) has therefore been a primary
focus of previous global convection simulations (Glatzmaier 1987;
Miesch et al.\ 2000; Elliott, Miesch \& Toomre 2000; Brun \& Toomre
2002).  Of particular importance is the mean longitudinal flow, i.e.\
the differential rotation, which is now reasonably well established from
helioseismic inversions, although investigations continue to
scrutinize its detailed spatial structure and temporal evolution
(Thompson et al. 2003).  The mean circulation in the meridional plane
has only been probed reliably in the surface layers of the sun through
Doppler measurements (Hathaway et al.\ 1996) and local-area
helioseismology (e.g.\ Haber et al.\ 2002).  Here we
discuss the mean flows achieved in our simulations and compare
them with solar observations and previous numerical models.

\subsection{Attributes of Mean Flows}

With fairly strong magnetic fields sustained within the bulk of the
convection zone in case {\em M3}, it is to be expected that the
differential rotation $\Omega$ will respond to the feedback from the
Lorentz forces. Figure 9 (left panel) shows the time-averaged angular
velocity achieved in case {\em M3}, which exhibits a prograde equatorial
rotation with a monotonic decrease in angular velocity toward higher
latitudes as in the sun.  The main effect of the Lorentz forces is to
extract energy from the differential rotation. The kinetic energy
contained in the differential rotation drops by a factor of two after
the addition of magnetic fields and this decrease accounts for over
70\% of the total kinetic energy difference (cf. \S 3.2).  This is
reflected by a 30\% decrease in the angular velocity contrast
$\Delta\Omega$ between the equator and latitudes of $60^\circ$, going from 140
nHz (or 34\% compared to the reference frame $\Omega_o$) in the
hydrodynamic case H to 100 nHz (or 24\%) in case {\em M3}.  This value
is close to the contrast of 22\% inferred from helioseismic inversion of
the solar profile (Thompson et al. 2003).  Thus the convection is
still able to maintain an almost solar-like angular velocity contrast
despite the inhibiting influence of Lorentz forces.

Eddy et al. (1976)
have deduced from a careful study of solar activity records 
during the Maunder minima that the sun was rotating about 3--4\% faster
in the equatorial region during that period than it does at present and
that the angular velocity contrast between the equator and latitudes
of $20^\circ$ may have been as much as a factor of three larger. The
somewhat faster rotation rate and larger $\Delta \Omega$ in case H 
(and {\em M2}) relative to case {\em M3} further suggests that a reduced 
level of the sun's magnetism may lead to greater differential
rotation (Brun 2004).

\begin{figure}[!ht]
\setlength{\unitlength}{1.0cm}
\begin{picture}(5,6.4)
\includegraphics{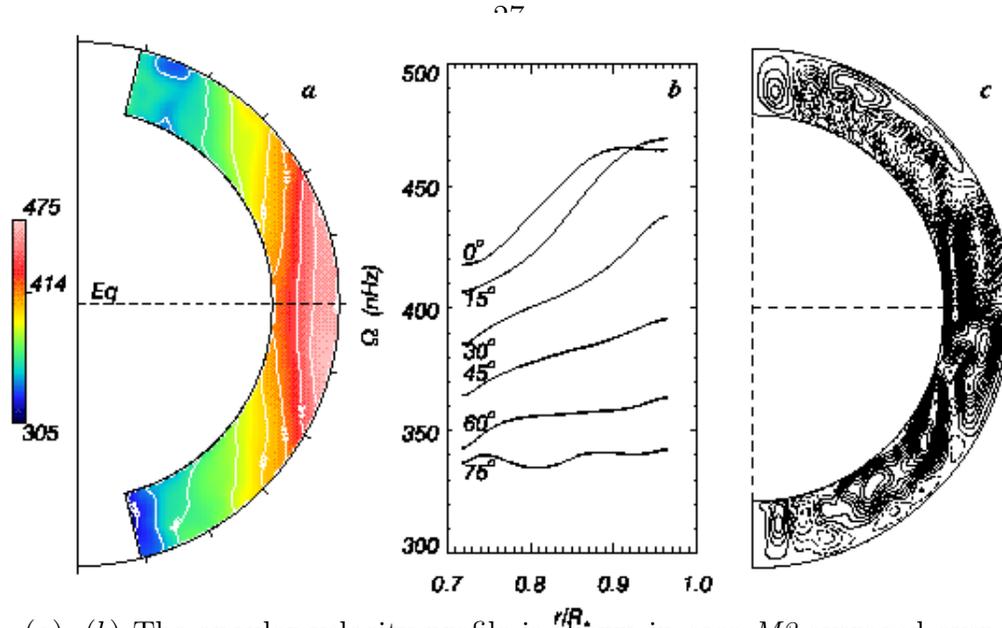}
\end{picture}
\caption[]{\label{fig9} ($a$), ($b$) The angular
velocity profile is shown in case {\em M3} averaged over longitude and time
(spanning an interval of 150 days late in the simulation).  
White/red and blue/green tones in
frame ($a$) denote faster and slower rotation respectively.  
Radial profiles are plotted in ($b$) for
selected latitudes.  ($c$) Displays the meridional circulation in
case {\em M3} averaged over longitude and time, represented as streamlines
of the mass flux.  Solid contours denote clockwise circulation and
dashed contours counter-clockwise circulation.}
\end{figure}

In Figure 9$c$ we display the meridional circulation realized in
case {\em M3}.  This meridional circulation is maintained by buoyancy
forces, Reynolds stresses, pressure gradients, Maxwell stresses, and
Coriolis forces acting on the differential rotation.  Since these
relatively large forces nearly cancel one another, the circulation can
be thought as a small departure from (magneto)-geostrophic balance,
and the presence of a magnetic field can clearly influence its subtle
maintenance.  In case {\em M3}, the meridional circulation exhibits a
multi-cell structure both in latitude and radius, and possesses some
asymmetry with respect to the equator. In particular, two vertical
cells are present at low latitudes in the northern hemisphere whereas
only one is present in the southern hemisphere.  Since the convection
possesses some asymmetry (cf. Fig.\ 4) it is not surprising that
the meridional circulation does the same.

Given the competing processes for its origin, this weak flow is not
straightforward to predict.  Typical amplitudes for the velocity are
of order 25 m s$^{-1}$, comparable to local helioseismic deductions
(Haber et al. 2002). The flow near the outer boundary is directed
poleward at low latitudes, with return flow deeper down. The temporal
fluctuations in the meridional circulation are large and thus stable
time averages are only attained by frequent sampling over many
rotation periods. The kinetic energy contained in the meridional
circulation (MCKE) is about two orders of magnitude smaller than that
contained in the differential rotation and convective motions and is
more than an order of magnitude less than the total magnetic energy
(ME; see Table 2). As a result, small fluctuations in the convective
motions, differential rotation and Lorentz forces can lead to major
variations in the circulation. Some of the helioseismic inferences
suggest the presence of single cell circulations which are at odds
with our multi-cell patterns.  However, these inferences vary from
year to year and there is recent evidence that multiple-cell structure
and equatorial asymmetries are developing in the meridional
circulation patterns just below the photosphere as the current solar
cycle advances (Haber et al. 2002).

\subsection{Redistribution of Angular Momentum}

We can better understand how the differential rotation profile in case
{\em M3} is achieved by identifying the main physical processes
responsible for redistributing angular momentum within our rotating
convective shells.  Our choice of stress-free and potential-field boundary 
conditions at the top and bottom of the computational domain have the 
advantage that no net external torque is applied, and thus angular momentum is 
conserved.  We may assess the transport of angular momentum within these
systems by considering the mean radial (${\cal F}_r$) and latitudinal
(${\cal F}_{\theta}$) angular momentum fluxes, extending the
procedure used in Brun \& Toomre (2002) to the magnetic context (see
also Elliott, Miesch \& Toomre 2000). Let us consider the
$\phi$-component of the momentum equation expressed in conservative
form and averaged in time and longitude:

\begin{equation}
\frac{1}{r^2} \frac{\p(r^2 {\cal F}_r)}{\p r}+\frac{1}{r \sin\theta}
\frac{\p(\sin \theta {\cal F}_{\theta})}{\p
\theta}=0,
\end{equation}
involving the mean  radial angular momentum flux
\begin{equation}
{\cal F}_r=\rb r\sin\theta[-\nu r\frac{\p}{\p
r}\left(\frac{\hat{v}}{r}\right)+\widehat{v_{r}^{'}
v_{\phi}^{'}}+\hat{v}_r(\hat{v}_{\phi}+\Omega r\sin\theta)-\frac{1}{4\pi\rb}\widehat{B_{r}^{'}
B_{\phi}^{'}}-\frac{1}{4\pi\rb}\hat{B}_r\hat{B}_{\phi}] \end{equation}
and the mean latitudinal angular momentum flux
\begin{equation}
{\cal F}_{\theta}=\rb r\sin\theta[-\nu
\frac{\sin\theta}{r}\frac{\p}{\p
\theta}\left(\frac{\hat{v}_{\phi}}{\sin\theta}\right)+\widehat
{v_{\theta}^{'} v_{\phi}^{'}}+\hat{v}_{\theta}(\hat{v}_{\phi}+\Omega
r\sin\theta)-\frac{1}{4\pi\rb}\widehat{B_{\theta}^{'}
B_{\phi}^{'}}-\frac{1}{4\pi\rb}\hat{B}_{\theta}\hat{B}_{\phi}].
\end{equation}

In the above expressions for both fluxes, the terms on the
right-hand-side denote contributions respectively from viscous
diffusion (which we denote as ${\cal F}_r^{VD}$ and ${\cal
F}_\theta^{VD}$), Reynolds stresses (${\cal F}_r^{RS}$ and ${\cal
F}_\theta^{RS}$), meridional circulation (${\cal F}_r^{MC}$ and ${\cal
F}_\theta^{MC}$), Maxwell stresses (${\cal F}_r^{MS}$ and ${\cal
F}_\theta^{MS}$) and large-scale Magnetic torques (${\cal F}_r^{MT}$
and ${\cal F}_\theta^{MT}$). The Reynolds stresses are associated with
correlations of the fluctuating velocity components which arise from
organized tilts within the convective structures, especially in the
downflow plumes (e.g. Brummell et al. 1998, Miesch et al. 2000).  In
the same spirit the Maxwell stresses are associated with correlations
of the fluctuating magnetic field components which arise from tilt
and twist within the magnetic structures.

\begin{figure}[!ht]
\setlength{\unitlength}{1.0cm}
\begin{picture}(5,6.8)
\includegraphics{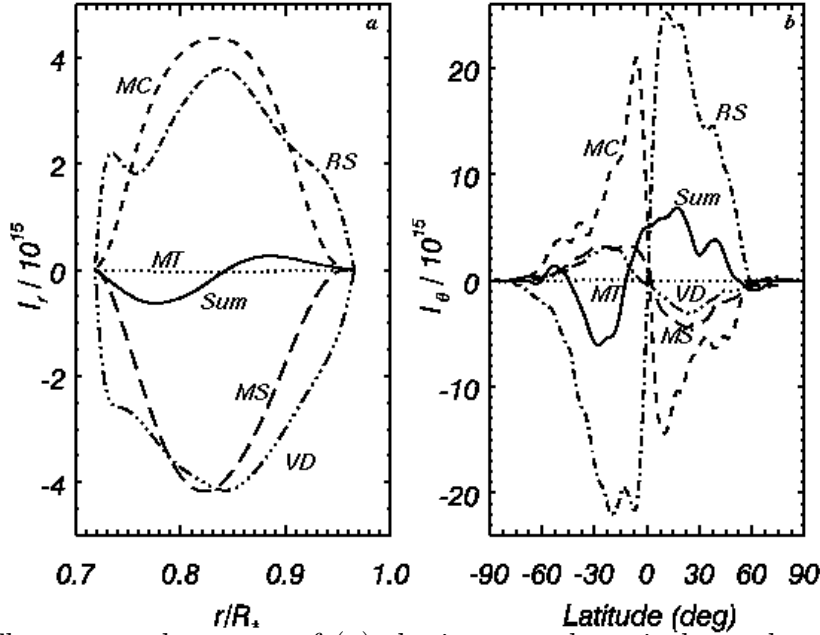}
\end{picture}
\caption[]{\label{amom} The temporal average of ($a$) the
integrated vertical angular momentum flux $I_r$ and ($b$) the integrated
latitudinal angular momentum flux $I_\theta$ for case {\em M3}.
The fluxes have been decomposed into their
viscous diffusion (labelled VD), Reynolds stress (RS), meridional
circulation (MC), Maxwell stress (MS) and large-scale magnetic torque
(MT) components.  The solid curves represent the sum of these
components and serve to indicate the quality of stationarity
achieved. Positive values represent a radial flux that is directed
outward, and a latitudinal flux directed from north to south. The
interval chosen for the time averages spans 150 days late in
the simulation (as in Fig. 9). The radial integrated flux $I_r$ 
has been normalized by $r_{top}^2$}
\end{figure}

In Figure 10 we show the components of  ${\cal F}_r$ and
${\cal F}_{\theta}$ for cases {\em M3}, having
integrated over co-latitude and radius as follows: 
\begin{equation}
I_r(r)=\int_0^{\pi} {\cal F}_r(r,\theta) \, r^2 \sin\theta
\, d\theta \; \mbox{ , } \; I_{\theta}(\theta)=\int_{r_{bot}}^{r_{top}} {\cal
F}_{\theta}(r,\theta) \, r \sin\theta \, dr \, ,
\end{equation}
Thus $I_r$ represents the net angular momentum flux through
horizontal shells at different radii and $I_\theta$ represents the net
flux through cones at different latitudes.  We then identify in turn
the contributions from viscous diffusion (VD), Reynolds stresses (RS),
meridional circulation (MC), Maxwell stresses (MS) and large scale
magnetic torques (MT). This representation is helpful in assessing
the sense and amplitude of angular momentum transport within
the convective shells by each component of ${\cal F}_r$ and 
${\cal F}_{\theta}$.

Turning first to Figure 10$a$, we see that the radial differential
rotation is being maintained by Reynolds stresses and meridional
circulation, $I_r^{RS}$ and $I_r^{MC}$, which both transport angular
momentum radially outward.  This outward transport is opposed by the
viscous flux $I_r^{VD}$, which is radially inward as implied by the
positive radial angular velocity gradient seen in Figure 9$b$.  The
Maxwell stresses $I_r^{MS}$ also act to oppose the generation of
differential rotation by the convection, possessing the same sign and
amplitude as the viscous torque.  The large-scale magnetic torques are
very small but negative as well, helping to decelerate the surface and
speed up the bottom of the shell.  The net radial flux $I_r$,
represented by the solid curve, is nearly zero, indicating that the
flow has achieved an approximate statistical equilibrium and that our
sampling in time captures this equilibrated state reasonably well,
despite the large temporal variations typically present in our
simulations.

The latitudinal angular momentum flux $I_{\theta}$ exhibits a more
complicated interplay among its various components than $I_r$, as
demonstrated in Figure 10$b$.  Here the angular momentum transport is
dominated by Reynolds stresses $I_\theta^{RS}$ which are consistently
directed toward the equator (i.e. negative in the southern hemisphere
and positive in the northern hemisphere).  This feature implies that
the equatorial acceleration observed in our simulations is mainly due
to the transport of angular momentum by the Reynolds stresses.
Further, unlike the radial angular momentum balance, we see that the
transport by meridional circulation $I_\theta^{MC}$ is opposite to
$F_\theta^{RS}$, with the meridional circulation seeking to slow down
the equator and speed up the poles. The viscous torque $I_\theta^{VD}$
is in the same direction but is a factor of four smaller in
amplitude. These results are identical to that deduced from case H
(much as in Brun \& Toomre 2002).  The main difference in case {\em
M3} comes from the Maxwell stress component $I_\theta^{MS}$, which
opposes the Reynolds stresses as in the radial angular momentum
balance.  The large-scale magnetic torque $I_\theta^{MT}$ is again
found to be negligible.  The total flux $I_{\theta}$ vacillates around
zero, indicating no net latitudinal angular momentum transport and an
acceptable equilibrated solution.

The reduction in the latitudinal contrast of $\Omega$ between cases
H and {\em M3} can be partially attributed to a global decrease in the
kinetic energy of the convection (see Table 2).  The rms Reynolds
number of case {\em M3} is about 12\% less than in case H, reflecting
the stablizing influence of magnetic fields.  However, the convection
kinetic energy is only reduced by about 27\% whereas the differential
rotation kinetic energy is reduced by over 50\%.  Figure 10
indicates that this large decrease in DRKE is due to the poleward
transport of angular momentum by Maxwell stresses. In case {\em M3} the
Reynolds stresses must balance the angular momentum transport by the
meridional circulation, the viscous diffusion, and the Maxwell
stresses, which leads to a less efficient acceleration of the
equatorial regions. Since the magnetic energy is only about 7\% of the
kinetic energy in case {\em M3} (cf. Table 2), the Maxwell stresses are
not the main players in redistributing the angular momentum, but
they do contribute more than the viscous torque $I_\theta^{VD}$ in the
latitudinal balance.  If the magnetic energy were to exceed about 20\% of the
total kinetic energy, Maxwell stresses and magnetic torques may become
strong enough to suppress the differential rotation almost entirely
(Gilman 1983; Brun 2004).  The sun may have ways of avoiding this
by expelling some of its magnetic flux.
  
We emphasize that the suppression of vertical and latitudinal
differential rotation by Lorentz forces in our simulations is
dominated by the fluctuating magnetic field components $I^{MS}$, not
the mean field components $I^{MT}$.  Magnetic tension forces
associated with the mean poloidal field do tend to inhibit rotational
shear as in axisymmetric models (MacGregor \& Charbonneau 1999), but
this intuitive ``rubber band'' effect is far less efficient than the
more subtle Maxwell stresses induced by correlations among the
turbulent magnetic field components.
  
Brun \& Toomre (2002) have found that as the level of turbulence is
increased, $I^{VD}$ reduces in amplitude and the transport of angular
momentum by the Reynolds stresses $I^{RS}$ and by the meridional
circulation $I^{MC}$ change accordingly to maintain equilibrium. Here
the presence of a fourth agent, namely the Maxwell stresses, can
modify this force balance and thus alter the equilibrium rotation
profile.

An important feature of the rotation profile in case H and also in
Case $AB$ of Brun \& Toomre (2002) is a monotonic decrease in angular
velocity with increasing latitude which persists all the way to the
polar regions.  This relatively slow polar rotation is supported by
helioseismic inversions but is generally difficult to achieve in
numerical simulations of convection because regions close to the
rotation axis undergo a prograde acceleration if fluid parcels tend to
conserve their angular momentum.  Thus, it is promising to see that
case {\em M3} has retained relatively slow rotation at high latitudes even
in the presence of magnetic fields.

Figure 10$b$ indicates that the the prograde equatorial rotation seen in
case {\em M3} is due to equatorward angular momentum transport by Reynolds
stresses and that the meridional circulation tends to oppose this
transport.  In many previous simulations, the poleward angular
momentum transport by the meridional circulation extends to higher
latitudes, tending to spin up the poles.  Thus the slow polar
rotation in case {\em M3} and its hydrodynamic predecessors, cases H and
$AB$, seems to come about from a relatively weak meridional
circulation at high latitudes (see also Brun \& Toomre 2002).  The
absence of strong high-latitude circulation cells permits a more
efficient extraction of angular momentum by the Reynolds stresses from
the polar regions toward the equator, yielding the interesting
differential rotation profile that is achieved. Since the Maxwell
stresses also transport angular momentum toward the poles, the polar
regions in case {\em M3} are found to rotate slightly faster than in case
H.  However, the angular momentum transport by Maxwell stresses is
distributed such that the global rotation retains the attribute of a
monotonic decrease of $\Omega$ with latitude.

\section{Evolution of Mean Magnetic Fields}\label{meanfields}

It is clear from the results presented (see e.g.\
Figs. 4, 7, 8, and 10) that the magnetic field is dominated by the
fluctuating or turbulent (non-axisymmetric) component.  However, the
mean (axisymmetric) field components have particular significance with
regard to solar dynamo theory, and thus it is instructive to explore their
structure and evolution in detail.  In particular, we wish to
understand our simulation results in the context of solar observations
although we are aware that we are still missing important ``dynamo
building blocks'' (c.f. \S1.1) such as magnetic pumping into a
tachocline-like shear layer.  Our results provide fundamental insight
into the generation of mean magnetic fields by turbulent convection
and as such can be used to evaluate and improve mean-field dynamo
models which do not explicitly consider the turbulent field and flow
components (e.g.\ Krause \& R\"adler 1980, Ossendrijver 2003).  In
what follows, we define the mean poloidal field in terms of the
longitudinally-averaged radial and latitudinal components,
$\left<B_p\right> = \left<B_r\right> \uvr + \left<B_\theta\right> \uvt$,
and the mean toroidal field in terms of 
the longitudinally-averaged longitudinal component 
$\left<B_t\right> = \left<B_\phi\right> \uvp$.

The generation of the mean toroidal field in our simulations is due 
to the shearing, stretching, and twisting of mean and fluctuating
poloidal fields by differential rotation (the $\omega$-effect)
and helical convective motions (the $\alpha$-effect).  Likewise, 
mean poloidal fields are generated from fluctuating toroidal fields 
via the $\alpha$-effect.  The $\alpha$-effect arises from correlations
between turbulent flows and fields as expressed in the mean 
(longitudinally-averaged) induction equation by the term 
$\Psi = \left<\curl\left({\bf v}^\prime \cross {\bf B}^\prime\right)\right>$,   
where primes indicate that the axisymmetric component has been 
subtracted off and angular brackets indicate a longitudinal average (Moffatt
1978, Stix 2002, Brandenburg \& Subramanian 2004). We find that the 
fluctuating fields in our simulations are much stronger 
than the mean fields, accounting for up to 98\% of the total magnetic 
energy, and the scale and amplitude of their correlations are not small 
in any sense and therefore cannot be reliably parameterized in terms 
of the mean field. It appears that the generation of mean fields in our 
simulations is not due to the $\alpha$-effect in the traditional sense, 
but rather to a more complex interplay between turbulent magnetic field 
and flow components. The chaotic nature of these turbulent components 
gives rise to intricate structure and aperiodic evolution in the mean fields.

\subsection{Poloidal Field}

Figure 11 illustrates the structure and evolution of the mean poloidal
field in case {\em M3}.  The top row shows four snapshots of the magnetic
lines of force of $\left<B_p\right>$ within the convective domain along with
a potential extrapolation of the external field up to 2 $R_*$.  The
initial seed field was dipolar (i.e antisymmetric with respect to the
equator), but symmetric fields (i.e quadrupolar configurations)
are also realized in our simulations, as in Figure 11$c$. The evolution
of the poloidal magnetic field from an antisymmetric to a symmetric
profile with respect to the equator is made possible because of the
nonlinear and asymmetric nature of the convection which amplifies the
field through dynamo action.  The continous exchange between dipolar
and quadrupolar topologies as well as higher-order multipoles results
in magnetic fields with intricate configurations and with no clear
equatorial symmetry preferences. Within the convective shell the
presence of strong magnetic field gradients and magnetic diffusion
lead to continous reconnection of the magnetic field lines.

\begin{figure}[!ht]
\setlength{\unitlength}{1.0cm}
\begin{picture}(5,9)
\includegraphics{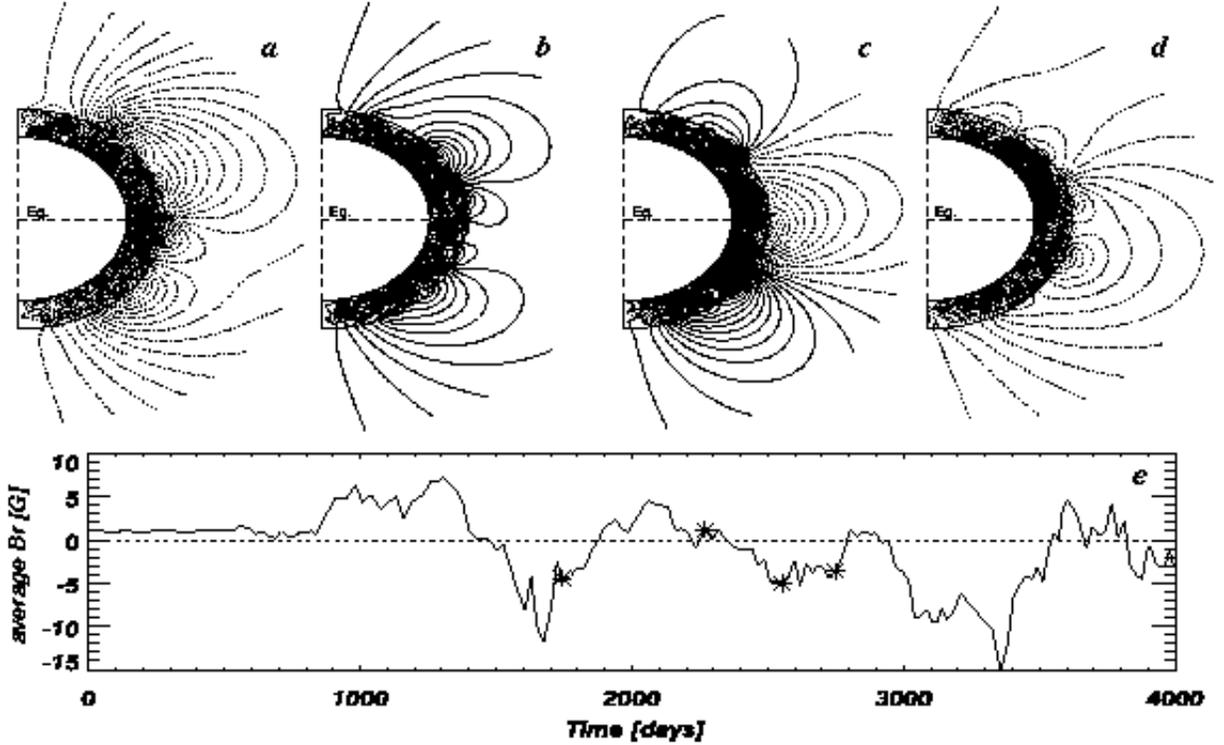}
\end{picture}
\caption[]{\label{Bpol} Temporal evolution of the mean poloidal
field is shown for case {\em M3}. ($a$)--($d$) The structure of the
field at four selected times after the magnetic energy has reached a
statistically steady state.  Solid contours denote positive polarity
(radially outward at the northern rotational pole) and dotted contours
denote negative polarity.  ($e$) The mean radial field at the outer
boundary averaged over the northern hemisphere, shown over the course
of the entire simulation.  The average polarity reverses after about
1750 days, and several more times afterward on a time scale of about
500 days.  However, the field generally exhibits a complex topology
with both symmetric as well as antisymmetric components.  The instants
in time corresponding to the upper frames are indicated in ($e$) with
asterisks.}
\end{figure}

The perpetual regeneration of magnetic flux by the convection can lead
to a global reversal of the magnetic field polarity.  Figure 11$e$
shows the temporal evolution of the average polarity of the poloidal
field in case {\em M3}, defined in terms of the radial magnetic field
$B_r$ averaged over the northern hemisphere of the outer boundary.
This is a measure of the total magnetic flux which passes through the
northern hemisphere at the outer surface of the shell, and since
$\nab\cdot {\bf B} = 0$ outside as well as inside the domain, the same
flux of opposite polarity must also pass through the southern
hemisphere.  A positive value indicates that the field is outward on
average in the northern hemisphere, as in the dipolar initial
conditions.  By contrast, a negative value indicates the average
polarity is opposite to that imposed in the initial conditions.

The flat evolution over the first 800 days corresponds to the linear
growth phase of the magnetic energy, where the field evolves slowly
away from its imposed initial dipolar topology and north-south
orientation. As the fields with negative polarity gain in strength, a
complex competition between the two polarities, directly related to
the turbulent nature of the dynamo, leads to a chaotic and irregular
variation of the average polarity.  Several field reversals do occur
on a time scale of about 500 days, but there is little evidence for
systematic cyclic behavior.  This time scale is comparable to the
1.5-year periods found by Gilman (1983) in some of his Boussinesq
dynamo simulations. Glatzmaier (1985a, 1987) inferred longer reversal
time scales ($\sim$ 10 years) in his simulations which incorporated
compressibility via the anelastic approximation as here and included
convective penetration into an underlying stable region.  Like
Glatzmaier's, our simulations only cover about 10 years so they would
not capture longer-term cyclic behavior if it were present.  However,
the chaotic short-term evolution suggests that longer-term periodic
behavior is unlikely for the configuration that we have adopted here.
  
Our high resolution simulations confirm that the time scale for field
reversal within the convective envelope itself is too short and that
without a stable layer such as the solar tachocline such simulations
are unlikely to reproduce the global-scale dynamo and 22-year activity
cycle observed in the sun. There is no systematic latitudinal
propagation of $\left<B_p\right>$ over the 4000 days that we have been able
to compute. Rather, the temporal evolution of $\left<B_p\right>$ is quite 
complex and highly unpredictable, governed by advection and amplification
by turbulent convective motions. Both Gilman and Glatzmaier found 
poleward propagation of $\left<B_p\right>$. The main difference
between their convective dynamo simulations and ours comes from the
level of turbulence and non-axisymmetry. In case {\em M3}, the
axisymmetric fields are weak and do not control the dynamical
evolution of the flow and magnetic fields, but seem on the contrary
passive, which could in part explain their erratic evolution.  
The mean poloidal field is generated mainly by the coupling
between fluctuating field and flow components and the generation
rate is not in general proportional to the strength of the mean
field as is assumed in the classical $\alpha$-effect. 

 
\subsection{Toroidal Fields}

\begin{figure}[!ht]
\setlength{\unitlength}{1.0cm}
\begin{picture}(5,6.5)
\includegraphics{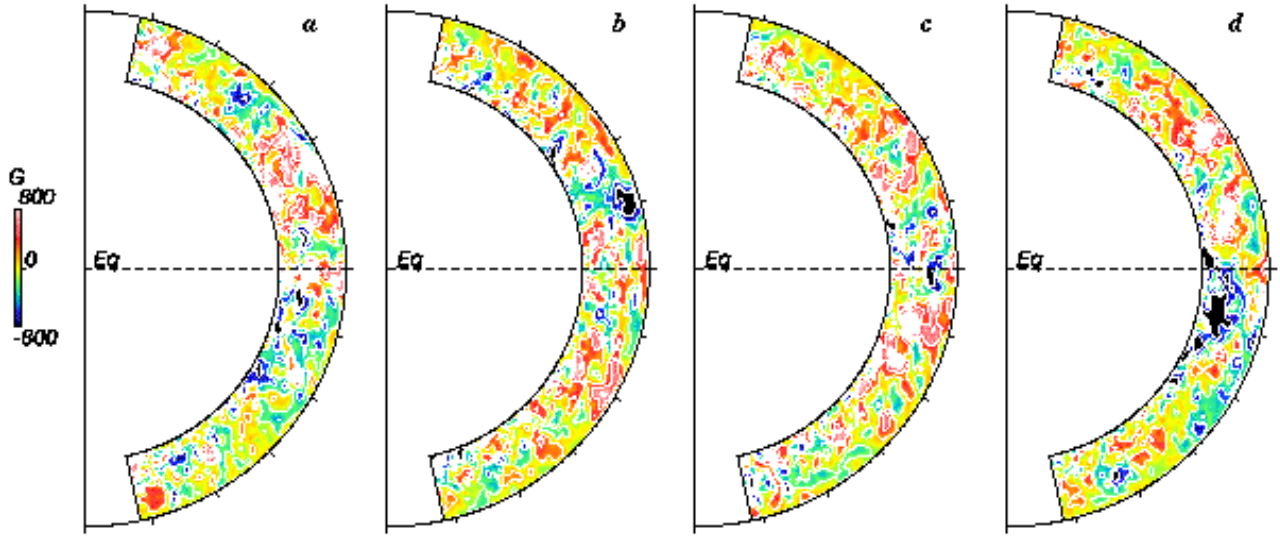}
\end{picture}
\caption[]{\label{Btor} The evolution of the mean toroidal field is shown for case {\em M3},
as a companion to Fig.\ 11.  Red and blue tones denote
eastward (prograde) and westward (retrograde) field respectively as indicated by
the color table.}
\end{figure}

Figure 12 shows the mean toroidal magnetic field
$\left<B_t\right>$ for the same time snapshots as displayed in Figure 11.  We
can readily see that it possesses small-scale structure which varies
substantially with time. Mixed polarities and intricate topologies are
present throughout the simulation domain, with no evident symmetries 
with respect to the equator. Instantaneous snapshots or time averages 
reveal weak responses with varying symmetries but these do not persist 
for any extended interval. Structures resembling thin tubes with
circular cross sections are present, but they generally do not remain
coherent long enough to rise and emerge through the surface due to
magnetic buoyancy.  No systematic latitudinal propagation of
$\left<B_t\right>$ is evident in this time sequence or in any that we have
studied. This is in contrast with solar observations which reveal
regular trends in the emergence of sunspots and related magnetic flux
over the 22-year activity cycle.

The mean toroidal field contains about 1.5\% of the total magnetic
energy, about a factor of three larger than the energy in the mean
poloidal field. The production terms in the mean induction equation 
due to differential rotation (the $\omega$-effect) and convective
motions (the $\alpha$-effect) are of the same order so our
simulations may be loosely classified as $\alpha^2-\omega$ dynamos.

In the sun the ratio of mean toroidal to poloidal magnetic energy is
at least 100, suggesting that the sun may not be generating
its mean toroidal field solely in the convective zone.  Glatzmaier
(1984; 1985a,b; 1987) incorporated convective penetration into an
underlying stable layer in his dynamo simulations and found that
$\left<B_t\right>$ was a significant fraction ($\sim 85\%$) of the
total magnetic energy. This result suggests that strong axisymmetric
toroidal fields are generated mainly in the stable layer via the
$\omega$-effect and strengthens the current paradigm that convection
in the solar envelope cannot amplify the mean toroidal field to
observed levels without the presence of convective penetration into a
stably-stratified shear layer such as the solar tachocline.  The 
convection zone continously supplies disorganized magnetic fields 
over a wide range of spatial scales to the tachocline, where they
are then amplified and organized into extended toroidal structures.


\section{Further Aspects of Field Generation}

\subsection{Helicity in Flows and Fields}\label{helicity}

It has long been realized that helicity can play an essential role in
hydromagnetic dynamo action, particularly in the solar context.
Parker's (1955) classical paradigm for the solar dynamo relies on
twisting motions in order to generate poloidal field from toroidal
field and thus drive the solar activity cycle.  Mean-field analyses of
homogeneous MHD turbulence based on the assumption of scale separation
yield an explicit expression for the regeneration rate of the 
magnetic field (the $\alpha$-effect) which is directly
proportional to the kinetic helicity of the flow, defined as the 
dot product of velocity and vorticity: 
$H_k = \vort \cdot {\bf v}$ (e.g.\ Moffatt 1978; Krause \& R\"adler 1980).
      
The kinetic helicity provides a measure of how much twist is
present in the velocity field.   Magnetic twist (and 
writhe, cf.\ Moffatt \& Ricca 1992) is often measured
by the magnetic helicity, defined as the dot product of
the magnetic field and the vector potential:
$H_m = {\bf A} \cdot {\bf B}$.   This quantity has particular
theoretical significance because it is conserved in ideal
(dissipationless) MHD (Biskamp 1993).   However, magnetic
helicity is very difficult to measure reliably on the sun.
From an observational standpoint, a more practical measure 
of magnetic twist is the current helicity, defined as the
scalar product of the magnetic field and current density:
$H_c = {\bf J} \cdot {\bf B}$.

Measurements of the radial component of the current helicity in the
solar photosphere have revealed a weak latitudinal dependence, tending
toward negative values in the northern hemisphere and positive values
in the southern hemisphere (Pevtsov, Canfield \& Metcalf 1994, 1995).
Helicity indicators in the chromosphere and corona reveal similar
hemisphere rules for a variety of structures; the pattern is
particularly strong for relatively large-scale features such as x-ray
sigmoids (Zirker et al.\ 1997; Pevtsov 2002; Pevtsov, Balasubramaniam,
\& Rogers 2003).  It has been suggested that the expulsion of this
magnetic helicity by coronal mass ejections may play a crucial role in
altering the global topology of the coronal field during polarity
reversals (Low 2001; Low \& Zhang 2004).

Figure 13 illustrates the kinetic and current
helicity in simulation {\em M3}.  The kinetic helicity shows
a clear variation with latitude.  Its amplitude peaks in the 
upper convection zone where it is negative in the northern
hemisphere and positive in the southern hemisphere, reflecting
the influence of rotation and density stratification; expanding
upflows spin down and contracting downflows spin up, tending
to conserve their angular momentum (e.g.\  Miesch et al.\ 2000).
In the lower convection zone the helicity reverses as downflows
encounter the lower boundary and diverge, inducing anticyclonic
vorticity.   The horizontal view in Figure 13
indicates that much of the kinetic helicity is confined to 
downflow lanes, reflecting their vortical nature.

\begin{figure}[!ht]
\setlength{\unitlength}{1.0cm}
\begin{picture}(5,11)
\includegraphics{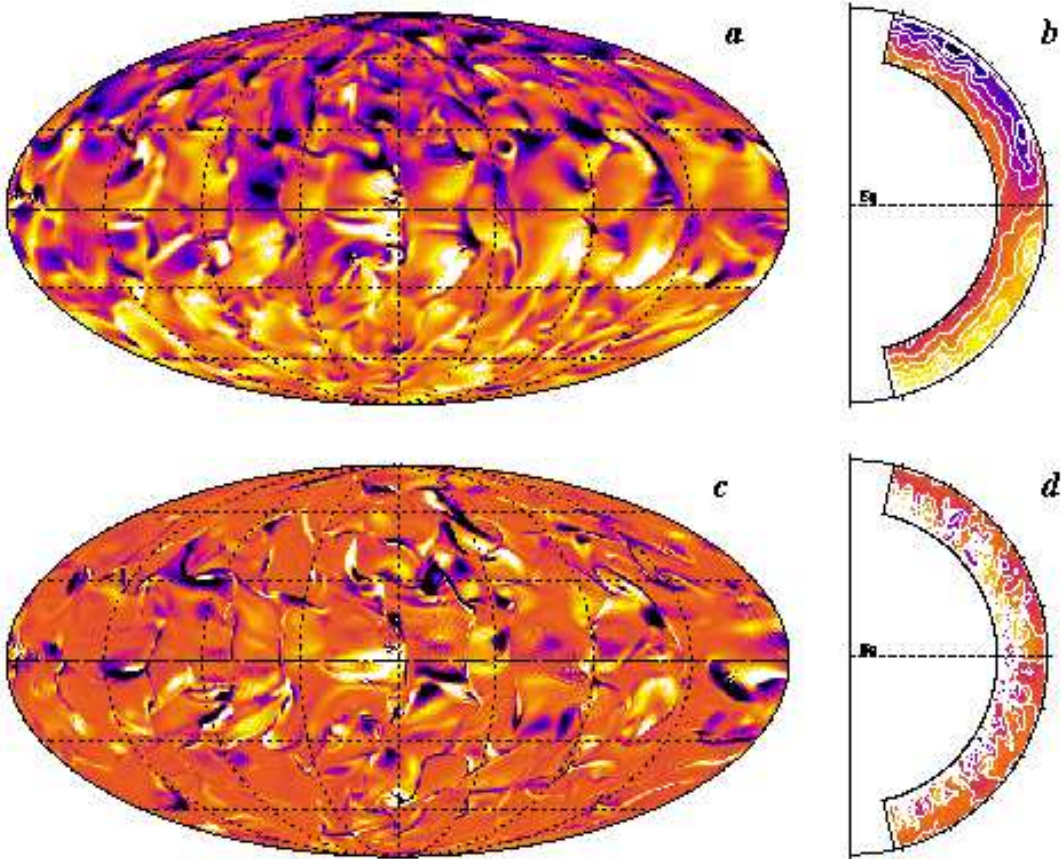}
\end{picture}
\caption[]{($a$, $b$) The kinetic helicity $H_k$ and ($c$, $d$) the current 
helicity $H_c$ in case {\em M3}.  The left column ($a$, $c$) shows
global views at the same time and horizontal level (near the top of the layer)
as in Fig.\ 4 and the right column ($b$, $d$) displays meridional
profiles averaged over longitude and time.  Bright tones denote
positive values and dark tones negative values (the color table is
as in Fig.\ 1).}
\end{figure}

The current helicity also tends to peak in downflow lanes but its latitudinal
variation is much less systematic than the kinetic helicity.  Current helicity
of both signs appears in each hemisphere, often juxtaposed in the same 
downflow lane.  The amplitude of the magnetic helicity peaks in the lower
convection zone where there is a weak pattern of positive and negative
values in the northern and southern hemisphere respectively.

These simulation results suggest that the helicity patterns observed in
the solar atmosphere may not be produced by turbulent convection in
the envelope.  Rather, they may originate in the tachocline where flux
tubes are formed and subsequently rise to the surface due to magnetic
buoyancy to form active regions.  Alternatively, the patterns may
arise from the action of Coriolis forces as flux tubes rise through
the convection zone or from footpoint motions after they
have emerged (e.g.\ Pevtsov 2002; Fan 2004).
 
\subsection{Spectral Distributions}

The convection patterns shown in Figure 4 suggest that the magnetic field
posesses relatively more small-scale structure than the velocity field.
This is verified by the energy spectra shown in Figure 14.  
The slope of the magnetic energy spectrum is much shallower than the 
kinetic energy spectrum and generally peaks at higher wavenumbers.
Thus the magnetic energy equals or exceeds the kinetic energy at small
scales, even though the ratio of total magnetic to kinetic energy remains
small.  Throughout most of the convection zone, the magnetic energy spectrum
peaks near $l \sim 30$, compared with $l \sim$ 12--15 for the kinetic
energy.   Near the top of the domain the kinetic energy spectrum peaks
at somewhat larger scales $l \sim 10$ whereas the magnetic energy 
spectrum remains relatively flat from $l = $ 1--20.   

\begin{figure}[!ht]
\setlength{\unitlength}{1.0cm}
\begin{picture}(5,4.5)
\includegraphics{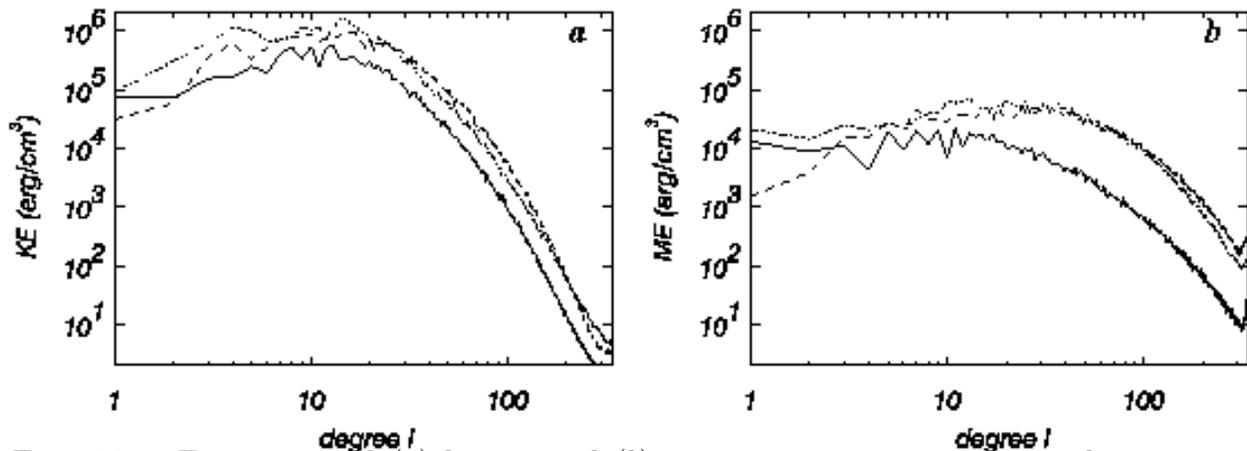}
\end{picture}
\caption[]{\label{spectra} Time-averaged ($a$) kinetic and ($b$) magnetic 
energy spectra are shown versus spherical harmonic degree $l$ 
(including all azimuthal wavenumbers $m$ but $m=0$), for case {\em M3} near 
the top, middle, and bottom of the convection zone 
(solid, dashed, and dotted lines respectively).}
\end{figure}

At degrees $l \gtrsim 30$, the spectra suggest some power-law behavior 
but it extends for less than a decade in degree so these simulations 
do not possess an extended inertial range.  The slope of the kinetic energy 
spectrum is substantially steeper than that expected for homogeneous,
isotropic, incompressible turbulence, with ($l^{-3/2}$) or without
($l^{-5/3}$) magnetic fields (e.g.\ Biskamp 1993).  Estimates based on
curve fits to the kinetic energy spectrum yield slopes steeper 
than $l^{-3}$.  The magnetic energy spectra are shallower but still
fall off faster than predicted for homogeneous, isotropic, incompressible 
MHD turbulence ($l^{-3/2}$).


\subsection{Probability Density Functions}\label{pdfsec}

Probability density functions (pdfs) can generally provide more
information about the structure and dynamics of a flow than spectral
analyses alone.  Indeed, in a homogeneous flow, the energy spectra
are simply related to the first moment of the corresponding 
two-point pdf.  We here consider the one-point pdf of 
the velocity and magnetic field variables as given by the histogram 
of values at all grid points, corrected for the grid convergence 
at the poles.  Results are shown in Figure 15.

\begin{figure}[!ht]
\setlength{\unitlength}{1.0cm}
\begin{picture}(5,7.5)
\includegraphics{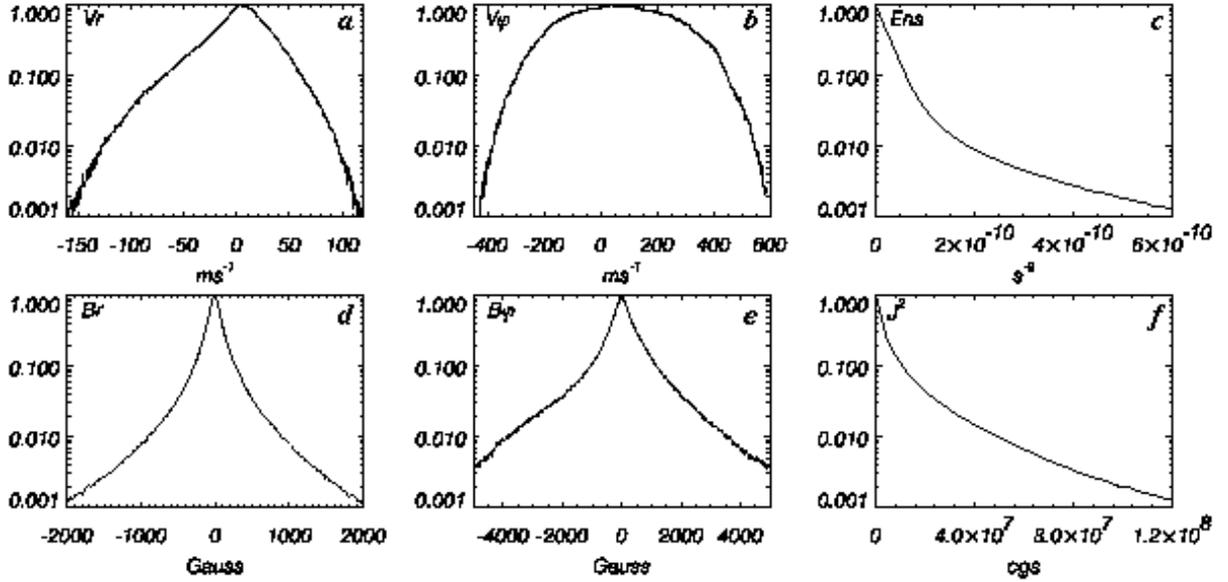} 
\end{picture}
\caption[]{\label{pdfs} Probablity density functions (pdfs) of 
($a$) the radial velocity $v_r$, ($b$) the zonal velocity $v_\phi$,
($c$) the enstrophy $\omega^2$, ($d$) the radial magnetic field
$B_r$, ($e$) the toroidal magnetic field $B_\phi$, and ($f$)
the square of the current density $J^2$ in case {\em M3}.  All pdfs are 
averaged over time and correspond to a horizontal level near 
the top of the shell.}
\end{figure}

Idealized isotropic, homogeneous turbulence has Gaussian velocity pdfs
but departures from Gaussian statistics are known to be present in
real-world turbulent flows.  Velocity differences and derivatives in
particular generally have non-Gaussian pdfs which are often
characterized by stretched exponentials $\exp[-\beta]$ with $0.5
\leq \beta \leq 2$ (e.g.\ She, Jackson \& Orszag 1988; Castaing, Gagne
\& Hopfinger 1990; Vincent \& Meneguzzi 1991; Kailasnath, Sreenivasan 
\& Stolovitzky 1992; Pumir 1996).  The tails of the distributions are 
often nearly exponential ($\beta \approx 1$) but can be even flatter, 
particularly in the viscous dissipation range. A flat slope ($\beta < 2$) 
indicates an excess of extreme (high-amplitude) events relative to 
a Gaussian distribution, thus reflecting spatial intermittency 
in the flow which may be associated with coherent structures 
(e.g., Vincent \& Meneguzzi 1991; Lamballais, Lesieur \& M\'etais 1997).

Another way to quantify the asymmetry and intermittency of selected 
flow and field variables is through moments of the pdf, in particular 
the skewness ${\cal S}$ and kurtosis ${\cal K}$, defined as:
\begin{equation}
{\cal S} = \frac{\int (x-\mu)^3 f(x) ~ dx}{\sigma^3 \int f(x) dx}  
~~~, \mbox{\hspace{1in}}
{\cal K} = \frac{\int (x-\mu)^4 f(x) ~ dx}{\sigma^4 \int f(x) dx} ~~~,  
\end{equation}
where $f(x)$ is the pdf, $x$ is the abscissa, $\mu$ is the mode of
the distribution, and $\sigma$ is the standard deviation:
\begin{equation}
\sigma = \left(\frac{\int (x-\mu)^2 f(x) ~ dx}{\int f(x) dx} \right)^{1/2} ~~~~.   
\end{equation}
Gaussian pdfs are characterized by ${\cal S} = 0$ and ${\cal K} = 3$
and exponential pdfs ($\beta = 1$) by ${\cal S} = 0$ and ${\cal K} = 6$.
A large value for ${\cal S}$ indicates asymmetry in the pdf whereas a
large value of ${\cal K}$ indicates a high degree of spatial
intermittency.

Probability density functions for turbulent, compressible, MHD convection
in Cartesian geometries have been reported by Brandenburg et al.\ (1996).
They found that the velocity pdfs were generally asymmetric and intermittent,
with ${\cal K} \simeq$ 4--5 for the horizontal components and 
${\cal K} \simeq 8$ for the vertical component.  The vorticity, 
magnetic field, and current density were more symmetric but also 
much more intermittent, possesing kurtosis values of ${\cal K} \simeq 20$
for ${\bf B}$ and ${\bf \omega}$ and ${\cal K} \simeq 30$ for ${\bf J}$.

The pdfs in case {\em M3} (Fig.\ 15) are qualitatively similar to those
found by Brandenburg et al.\ (1996).  The radial velocity has nearly
exponential tails (${\cal K}=4.6$) and a negative skewness (${\cal
S}=-0.98$); the fastest downflows are more than 150 m s$^{-1}$
compared to about 120 m s$^{-1}$ for upflows.  The zonal velocity,
$v_\phi$, is more Gaussian (${\cal K}=2.4$) but still asymmetric
(${\cal S}=0.45$), reflecting the influence of the differential
rotation.  The enstrophy also appears nearly exponential but with two
distinct slopes, flatting out for the highest-amplitude events.  This
implies a very high degree of intermittency (${\cal K}=270$).  The
radial and toroidal magnetic fields are more intermittent than the
velocity field (${\cal K}=$ 79, 11) and they appear to be more
symmetric, although several outlier points in the extreme tails of the
$B_r$ pdf give rise to a relatively large skewness, ${\cal S}=-1.3$
(${\cal S}=0.18$ for the $B_\phi$ pdf).  Maximum field strengths reach
about 5000 G for the toroidal field and somewhat less (2000 G) for the
radial field.  The relatively intermittent spatial structure of the
magnetic field is also apparent in the pdf of $J^2$, which possesses
an even higher kurtosis (${\cal K}=440$) than the enstrophy.  Note
that the kurtosis values quoted here for the enstrophy and current
density pdfs are much higher than those reported by Brandenburg et
al.\ (1996) primarily because they considered the linear vector fields 
${\bf \omega}$ and ${\bf J}$ whereas we have considered the nonlinear
scalar products $\omega^2$ and $J^2$. 

\section{Summary}

In this paper we report the highest-resolution 3--D simulations
achieved to date of hydromagnetic dynamo action by global-scale
turbulent convection and differential rotation in the solar envelope.
Building upon our own previous hydrodynamic simulations and the
pioneering dynamo models of Gilman (1983) and Glatzmaier (1984), we
have investigated the generation and maintenance of mean and
fluctuating magnetic fields in the solar convection zone, focusing on
their structure and evolution and on their dynamical influence upon the
flow field through Lorentz forces.  Our simulations are not intended
to provide a comprehensive model of the solar dynamo; they do not
address important ingredients such as toroidal field generation and
storage in the tachocline or flux emergence through the photosphere.
Still, they provide essential insight into a crucial element of the
global dynamo process, namely the generation of and coupling between
poloidal and toroidal magnetic fields in the convection zone, which 
is often described in terms of an $\alpha$-effect and $\omega$-effect.

The numerical experiments we have performed involve the addition of a
small seed magnetic field to an existing hydrodynamic simulation.  For
the parameter regimes considered here, we find that sustained dynamo
action occurs when the magnetic Reynolds number $R_m$ exceeds
about 300.  If this is the case, then the seed magnetic field grows
exponentially and subsequently saturates, reaching a statistically steady
state as Lorentz forces begin to feed back on the flow field.
Throughout most of this paper we focus on our simulation with the
lowest magnetic diffusivity, case {\em M3} ($R_m = 486$), in which
the steady-state magnetic energy is about 7\% of the total kinetic
energy contained in the convection and differential rotation.  At this
level of magnetism, Lorentz forces are not strong enough to
dramatically change the appearance of the flow; convective patterns in
case {\em M3} are similar to those in the non-magnetic progenitor
simulation, case H.  Furthermore, the radial energy flux balance through
the shell is essentially unaffected; the Poynting flux is neglible and
the net convective and diffusive energy fluxes are nearly the same as
in case H.

Although Lorentz forces have little effect on the appearance of the
convection in case {\em M3} relative to case H, they do have a
substantial influence on the structure and evolution of mean flows.
Fluctuating magnetic fields transport angular momentum poleward via
Maxwell stresses, decreasing the magntitude of the differential
rotation.  This leads to a decrease in the angular velocity contrast
between the equator and latitudes of $60^\circ$ from 34\% in case H to
24\% in case {\em M3}.  Magnetic tension forces associated with the mean
(axisymmetric) poloidal field also tend to suppress rotational shear
but this process is much less efficient than Maxwell stresses and
plays a negligible role in the maintenance of the global rotation
profile.  Despite the inhibiting effects of Lorentz forces, case {\em M3}
is able to sustain a strong differential rotation comparable in
amplitude and structure to the solar internal rotation inferred from
helioseismology.

The meridional circulation in the simulations reported here is
generally characterized by multiple cells in latitude and radius and
large temporal variations over time scales of weeks and months.  This
spatial and temporal variation is not surprising, since the
circulation arises from small differences in large forces which
fluctuate substantially in space and time, including Reynolds
stresses, Maxwell stresses, thermal (baroclinic) driving, and Coriolis
forces operating on the differential rotation.  Still, it is in sharp
contrast to many solar dynamo models which assume that the meridional
circulation is relatively smooth and steady, dominated by one or two
large cells in each hemisphere (e.g.\ Wang \& Sheeley 1991; Durney
1997; Dikpati \& Charbonneau 1999).  Doppler measurements of
photospheric flows and local-domain helioseismic inversions typically
reveal systematic circulation patterns in the surface layers of the
sun, with poleward flow of about 20 m s$^{-1}$ at low latitudes and
some time variation (e.g.\ Hathaway et al.\ 1996; Haber et al.\ 2000).
Although these analyses lie outside our computational domain, the
results are roughly consistent with our simulations: the meridional
flows at low latitudes near the top of the shell are consistently
poleward when averaged over several months, although these cells only
extend up to about 30$^\circ$ in latitude (see Fig.\ 9).  Somewhat
deeper helioseismic probing (down to $\sim$ 0.98 $R_*$), still in
local domains, provides some evidence for large temporal fluctuations
and multiple cells in radius (Haber et al.\ 2002) but little is
currently known about circulation patterns below about 0.97 $R_*$.
Characterizing the nature of the meridional flow in the deep
convection zone is thus an important future challenge for
helioseismology.

About 98\% of the magnetic energy in our simulations is contained in
the fluctuating (non-axisymmetric) field components which dominate the
Lorentz forces and the induction equation.  These components
exhibit a complex spatial structure and rapid time evolution as they
are amplified, advected, and distorted by convective motions.  The
distortion is particularly pronounced at mid latitudes, around 
$25^\circ$, where there is a change in the nature of the convective
patterns from the north-south aligned downflow structures which
dominate the equatorial regions to the more isotropic high-latitude
network.  The magnetic field possesses more small-scale structure and
is significantly more intermittent than the velocity field, a result
which is best demonstrated by considering the spectra and probability
density functions (pdfs) presented in \S7. The pdfs are in general
non-Gaussian and asymmetric, in contrast to homogeneous, isotropic
turbulence.

There is a noticable difference in the topology of the radial and
toroidal magnetic field components. Particularly near the top of the
convection zone, the radial field $B_r$ concentrates in downflow
lanes where fields of opposite polarity are brought together by
converging horizontal flows, thus promoting magnetic reconnection, and
where magnetic structures are twisted and distorted by vorticity and
shear.  Both the kinetic helicity $\mbox{\boldmath $\omega$} {\bf
\cdot v}$, and the current helicity ${\bf J \cdot B}$ peak in these
vortical downflow lanes.  However, unlike the kinetic helicity, the
current helicity does not exhibit a strong sign reversal in the
northern and southern hemispheres; both signs are distributed across
all latitudes, often in close proximity.  A potential extrapolation of
the radial field beyond the outer boundary of our domain reveals a
complex web of magnetic loops, exhibiting both local and long-range
connectivity across the surface.

Relative to the radial field, the toroidal field $B_\phi$ is
organized into larger-scale ribbons and sheets which are stretched out
in longitude by the differential rotation and which are not in general
confined to downflow lanes.  Near the top of the convection zone,
broad patches of like-signed toroidal field exist between the
north-south aligned downflow lanes at low latitudes.  These patches
are typically confined to the surface layers, with a relatively small
radial extent.  Although some structures resemble toroidal flux tubes,
they are rapidly advected and distorted by convective motions and they
generally lose their identity before magnetic buoyancy forces would
otherwise cause them to rise and emerge from the surface.  Peak field
strengths reach about 4000--5000 G for the toroidal field and about
2000 G for the radial field.

The mean poloidal and toroidal fields have much smaller amplitudes
than the fluctuating fields in our simulations, but nevertheless they
have particular significance for solar dynamo theory.  Our simulations
do not exhibit the organized structure, systematic propagation
patterns, and periodic polarity reversals which are known to exist in
the sun.  Rather, they possess a relatively complex spatial and
temporal dependence which can be attributed to the highly nonlinear
nature of the fluctuating velocity and magnetic field correlations
through which they are principally maintained.  

The energy in the mean toroidal field in case {\em M3} is about a
factor of three larger than that in the mean poloidal field.  This
asymmetry suggests that the differential rotation plays an important
role in the generation of mean fields via the $\omega$-effect, in
addition to the contribution from convective motions which can be
loosely regarded as a (non-traditional) $\alpha$-effect.  By contrast,
the fluctuating poloidal and toroidal fields are comparable in
amplitude, suggesting that the $\omega$-effect plays a smaller role.
However, the influence of differential rotation is still evident in
the morphology of the fluctuating toroidal field (see Fig.\ 7$b$).
The magnitude of the mean poloidal field near the surface in our
simulations ($\sim$ 5--10G, see Fig.\ 11) is comparable to the
large-scale poloidal field at the surface of the sun estimated from
photospheric and coronal observations (e.g.\ Gibson et al.\ 1999).
However, the peak strength of the mean toroidal field in our
simulations ($\sim 800 G$) is less than the field strength thought to
exist in concentrated flux tubes in the solar interior where the
estimated field strength ranges from 10$^4$--10$^5$ G near the base of
the convection zone to $\sim$ 10$^3$ G near the surface (e.g.\ Fisher
et al.\ 2000).  The relatively weak toroidal fields in our simulations
can likely be attributed to the absence of a tachocline where toroidal
flux can be efficiently stored and amplified by strong rotational
shear. 

The tachocline is an essential ingredient of the solar dynamo which is
missing from the models reported here.  It likely plays a central role
in many aspects of the solar activity cycle, including the structure,
strength, and emergence latitudes of sunspots and active regions as
reflected for example by the well-known ``butterfly diagram''.  The
large differential rotation and stable stratification in the lower
portion of the tachocline promote the generation and storage of strong
toroidal fields which are thought to account for much of the magnetic
activity observed in the solar atmosphere.  Coupling between the
convection zone and the radiative interior may also help to regularize
the structure and evolution of the mean poloidal field, producing
dipole configurations and cyclic reversals.  We are now working to
improve our dynamo simulations by incorporating convective penetration
into an underlying stable region and a layer of rotational shear
similar in nature to the solar tachocline.  Results from these models
will be published in forthcoming papers.

\acknowledgments
We thank Nicholas Brummell, 
Marc DeRosa, Emmanuel Dormy, Peter Gilman, 
Annick Pouquet and Jean-Paul Zahn for helpful discussions.  
This work was partly supported by NASA through
SEC Theory Program grants NAG5-8133, NAG5-12815 and work orders W-10, 175
and W-10, 177, and by NSF through grant ATM-9731676. 
Various phases of the simulations with ASH were carried out with NSF PACI 
support of the San Diego Supercomputer Center (SDSC), the National Center 
for Supercomputing Applications (NCSA), and the Pittsburgh Supercomputing 
Center (PSC) as well as with the Centre de Calcul pour la Recherche et 
la Technologie (CCRT) of CEA at Bruy\`ere-le-Chatel.  Much of the analysis 
of the extensive data sets was carried out in the Laboratory for 
Computational Dynamics (LCD) within JILA.

\appendix
{\bf Appendix A: Model Equations}

The anelastic equations (1)--(7) (\S2.1) define our physical model.
Here we express these equations as they are solved by our numerical
algorithm, making use of the velocity and magnetic field decomposition
in equations (8)-(9).  Diagnostic equations for the streamfunctions and
potentials $W$, $Z$, $A$, and $C$ are obtained by considering the
vertical component of the momentum and induction equations and the
vertical component of their curl.  A Poisson equation for pressure can
then be derived by taking the divergence of the momentum equation.
However, the additional radial derivative this would require can
compromise the accuracy of the solution, particularly when applied to
the nonlinear advection terms.  Since horizontal derivatives are more
accurate than vertical derivatives, we choose to only take the
horizontal divergence of the momentum equations rather than the full
divergence.  This results in a diagnostic equation for the horizontal
divergence of the velocity field, which is proporational to $\pd W/\pd
r$.  A spherical harmonic transformation is applied to the governing
equations before they are discretized in time so the time stepping
occurs in spectral space: $(\ell,m,r)$.  After some manipulation, the
governing equations for the spherical harmonic coefficients of the
state variables can be expressed as follows:

\begin{eqnarray}
\label{W_eqn_ss}
\frac{\ell(\ell+1)}{r^2}
\frac{\pd W_{\ell m}}{\pd t} & = & {\cal L}^W + {\cal N}^W  
\\ \label{P_eqn_ss}
- \frac{\ell(\ell+1)}{r^2}
\frac{\pd}{\pd t} \left( \frac{\pd W_{\ell m}}{\pd r} \right) & = &
{\cal L}^P + {\cal N}^P 
\\ \label{Z_eqn_ss}
\frac{\ell(\ell+1)}{r^2}
\frac{\pd Z_{\ell m}}{\pd t} & = & {\cal L}^Z + {\cal N}^Z  
\\ \label{S_eqn_ss}
\frac{\pd S_{\ell m}}{\pd t} & = & {\cal L}^S + {\cal N}^S .
\\ \label{A_eqn_ss}
\frac{\ell(\ell+1)}{r^2}
\frac{\pd A_{\ell m}}{\pd t} & = & {\cal L}^A + {\cal N}^A .
\\ \label{C_eqn_ss}
\frac{\ell(\ell+1)}{r^2}
\frac{\pd C_{\ell m}}{\pd t} & = & {\cal L}^C + {\cal N}^C .
\end{eqnarray}

In these expressions, the ${\cal L}$ denote the linear diffusion, 
pressure gradient, buoyancy, and volume heating terms which are
implemented using a semi-implicit, Crank-Nicolson timestepping
method:
\begin{multline}
{\cal L}^W = 
- \frac{\pd P_{\ell m}}{\pd r} - g \rho_{\ell m}
+ \nu \left(\frac{\ell(\ell+1)}{r^2}\right) \left\{ 
\frac{\pd^2 W_{\ell m}}{\pd r^2} + \left( 2\frac{d \ln \nu}{dr}
- \frac{1}{3}\frac{d \ln \rb}{dr} \right)\frac{\pd W_{\ell m}}{\pd r}
\right. \\ - \left. \left[ \frac{4}{3} \left(
\frac{d \ln \nu}{dr}\frac{d \ln \rb}{dr} + \frac{d^2 \ln \rb}{dr^2}
+\frac{1}{r}\frac{d \ln \rb}{dr} + \frac{3}{r}\frac{d \ln \nu}{dr}
\right) + \frac{\ell(\ell+1)}{r^2}\right] W_{\ell m} \right\} \; , \nonumber 
\end{multline}
\begin{multline}
{\cal L}^P = \left(\frac{\ell(\ell + 1)}{r^2}\right) P_{l m}
- \nu \left(\frac{\ell(\ell+1)}{r^2}\right) \left\{ 
\frac{\pd^3 W_{\ell m}}{\pd r^3} + \left( \frac{d \ln \nu}{dr}
- \frac{d \ln \rb}{dr} \right) \frac{\pd^2 W_{\ell m}}{\pd r^2}
\right. \\ \left.
- \left( \frac{2}{r}\frac{d \ln \rb}{dr} + \frac{d^2 \ln \rb}{dr^2}
+\frac{2}{r}\frac{d \ln \nu}{dr}+\frac{d \ln \nu}{dr}\frac{d \ln \rb}{dr}
+ \frac{\ell(\ell+1)}{r^2} \right)\frac{\pd W_{\ell m}}{\pd r}
\right. \\ \left.
- \left( \frac{d \ln \nu}{dr}+\frac{2}{r}+\frac{2}{3}\frac{d \ln \rb}{dr}
\right)\frac{\ell(\ell+1)}{r^2} W_{\ell m}
\right\} \; , \nonumber
\end{multline}
\begin{multline}
{\cal L}^Z =
\nu \left(\frac{\ell(\ell+1)}{r^2}\right) \left\{ 
\frac{\pd^2 Z_{\ell m}}{\pd r^2}
+ \left( \frac{d \ln \nu}{dr} - \frac{d \ln \rb}{dr} \right)
\frac{\pd Z_{\ell m}}{\pd r}
\right. \\ \left. 
- \left(\frac{2}{r}\frac{d \ln \nu}{dr} + \frac{d \ln \rb}{dr}\;
\frac{d \ln \nu}{dr} + \frac{d^2 \ln \rb}{dr^2} + \frac{2}{r}
\frac{d \ln \rb}{dr} - \frac{\ell(\ell+1)}{r^2} \right) Z_{\ell m}  
\right\} \; , \nonumber
\end{multline}
\begin{multline}
{\cal L}^S =  - \frac{\ell(\ell+1)}{r^2} \frac{d\Sh}{dr} W_{\ell m} +  
\frac{\kappa_r C_P}{\tb} \left[ \frac{\pd^2}{\pd r^2}
+ \left( \frac{d}{dr}\ln(r^2\kappa_r\rb)\right)
\frac{\pd}{\pd r} \right] (T_{\ell m} + \tb)
- \frac{\kappa_r C_P}{\tb} \frac{\ell (\ell+1)}{r^2} T_{\ell m} 
\nonumber \\ + \kappa \left[ \frac{\pd^2}{\pd r^2}
+ \left( \frac{d}{dr} \ln (r^2 \kappa \rb \tb) \right)
\frac{\pd}{\pd r}\right] (S_{\ell m} + \Sh) 
- \kappa \frac{\ell (\ell + 1)}{r^2} S_{\ell m} \; ,\nonumber
\end{multline}
\begin{equation}
{\cal L}^A =  \eta \left(\frac{\ell(\ell+1)}{r^2}\right) \left\{ 
\frac{\pd^2 A_{\ell m}}{\pd r^2} + \frac{d \ln \eta}{dr}
\frac{\pd A_{\ell m}}{\pd r} - \frac{\ell(\ell+1)}{r^2} A_{\ell m} \right\} \; ,\nonumber
\end{equation}
and
\begin{equation}
{\cal L}^C =  \eta \left(\frac{\ell(\ell+1)}{r^2}\right) \left\{ 
\frac{\pd^2 C_{\ell m}}{\pd r^2} - \frac{\ell(\ell+1)}{r^2} C_{\ell m}\right\} \; .\nonumber
\end{equation}
The perfect gas equation of state implies
\begin{equation}
\rho_{\ell m} = \rh \left( 
\frac{1}{\gamma}\frac{P_{\ell m}}{\ph} - \frac{S_{\ell m}}{c_p} \right)
\end{equation}
and
\begin{equation}
T_{\ell m} = \Th \left( \frac{\gamma - 1}{\gamma}\frac{P_{\ell m}}{\ph}
+ \frac{S_{\ell m}}{c_p} \right) .
\end{equation}

The ${\cal N}$ terms in equations (A1)--(A6) include nonlinear advection
terms which are implemented using an explicit, two-level
Adams-Bashforth time stepping method.  Although the Coriolis terms are
formally linear, they are also included in the ${\cal N}$ terms
because, unlike the other linear terms, the resulting coefficients
depend on azimuthal wavenumber $m$, and they couple the vertical
vorticity equation to the vertical momentum and horizontal divergence
equations.  This would greatly complicate the matrix solution involved
in the Crank-Nicholson method.  Thus the ${\cal N}$ terms in the
momentum equations include Coriolis terms which can be written 
in spherical harmonic space as:
\begin{equation}
{\cal N}^W = {\cal A}_{\ell m}^W + \Lambda + \frac{2 \Omega_o}{r} 
   \left( \imath m \frac{\pd W_{\ell m}}{\pd r} -  
   (\ell-1) c_{\ell}^m Z_{\ell-1}^m + (\ell+2)c_{\ell+1}^m Z_{\ell+1}^m
   \right) \; , \nonumber
\end{equation}
\begin{equation}
{\cal N}^P = {\cal A}_{\ell m}^P  
   + \frac{2 \Omega_o}{r^2}
\left[ -\imath m \left( \frac{\pd W_{\ell m}}{\pd r}
   + \frac{\ell(\ell+1)}{r} W_{\ell m} \right) + (\ell^2-1) c_{\ell}^m
   Z_{\ell-1}^m + \ell (\ell+2) c_{\ell+1}^m Z_{\ell+1}^m \right] \; \nonumber , 
\end{equation}
and
\begin{multline}
{\cal N}^Z = {\cal A}_{\ell m}^Z  
   + \frac{2 \Omega_o}{r^2} \left( - \frac{\ell(\ell^2-1)}{r} c_{\ell}^m
   W_{\ell-1}^m + \frac{\ell(\ell+1)(\ell+2)}{r} c_{\ell+1}^m W_{\ell+1}^m 
   \right. \\ \left. 
   + (\ell^2-1) c_{\ell}^m \frac{\pd W_{\ell-1}^m}{\pd r}
   + \ell(\ell+2) c_{\ell+1}^m \frac{\pd W_{\ell+1}^m}{\pd r}
   + \imath m Z_{\ell m} \right) . \nonumber
\end{multline}
The ${\cal A}_{\ell m}^i$ in these equations represent the spherical harmonic 
coefficients of the nonlinear velocity advection terms and Lorentz forces.  
If we define their corresponding configuration space representation 
as:
\begin{equation}
{\cal A}^i(r,\theta,\phi,t) = \sum_{\ell, m} {\cal A}^i_{\ell m}(r,t) 
\; Y_{\ell m}(\theta,\phi) \mbox{\hspace{.5in}} [i=W,P,Z] \; ,
\end{equation}
then
\begin{equation}
{\cal A}^W = - \rb
\left( v_r\frac{\pd v_r}{\pd r} + \frac{v_\theta}{r}\frac{\pd v_r}{\pd \theta}
+ \frac{v_\phi}{r \sin \theta}\frac{\pd v_r}{\pd \phi} - 
\frac{v_\theta^2 + v_\phi^2}{r} \right) 
+ J_\theta B_\phi - J_\phi B_\theta \; , 
\end{equation}
\begin{equation}
{\cal A}^P = \frac{1}{r \sin\theta} \left\{
\frac{\pd}{\pd \theta} \left( \sin\theta {\cal A}_\theta \right)
+ \frac{\pd {\cal A}_\phi}{\pd \phi} 
\right\} \; , 
\end{equation}
and
\begin{equation}
A^Z = \frac{1}{r \sin\theta} \left\{
\frac{\pd}{\pd \theta} \left( \sin\theta {\cal A}_\phi \right)
- \frac{\pd {\cal A}_\theta}{\pd \phi} 
\right\} \; ,
\end{equation}
where 
\begin{equation}
{\cal A}_\theta = - \rb \left( v_r\frac{\pd v_\theta}{\pd r}
+ \frac{v_\theta}{r}\frac{\pd v_\theta}{\pd \theta}+\frac{v_\phi}{r \sin\theta}
\frac{\pd v_\theta}{\pd \phi} + \frac{v_r v_\theta}{r} - 
\frac{\cos\theta}{r \sin\theta}v_\phi^2 \right) + J_\phi B_r - J_r B_\phi ~~~,
\end{equation}
\begin{equation}
{\cal A}_\phi = - \rb \left( v_r\frac{\pd v_\phi}{\pd r} + \frac{v_\theta}{r}
\frac{\pd v_\phi}{\pd \theta} + \frac{v_\phi}{r \sin\theta}
\frac{\pd v_\phi}{\pd \phi} + \frac{v_r v_\phi}{r} + 
\frac{\cos\theta}{r \sin\theta} v_\theta v_\phi \right)  
+ J_r B_\theta - J_\theta B_r  ~~~,
\end{equation}
and $\JJ = \curl \BB / (4 \pi)$.  The dimensonal current density is
given by ${\bf j} = c \JJ$.

Likewise, the remaining ${\cal N}$ terms represent the spherical 
harmonic coefficients corresponding to the nonlinear terms in the 
energy and induction equations:
\begin{equation}
{\cal A}^i(r,\theta,\phi,t) = \sum_{\ell, m} {\cal N}^i(\ell,m,r,t) 
\; Y_{\ell m}(\theta,\phi) \mbox{\hspace{.5in}} [i=S,A,C] \; ,
\end{equation}
where
\begin{equation}
{\cal A}^S = v_r\frac{\pd s}{\pd r} + \frac{v_\theta}{r}\frac{\pd s}{\pd \theta}
+ \frac{v_\phi}{r \sin \theta}\frac{\pd s}{\pd \phi} 
         + \frac{2\nu}{\tb} \left\{ e_{ij} e_{ij} - \frac{1}{3} 
\left( v_r \frac{d \ln \rb}{d r}\right)^2 
\right\}  + \frac{4 \pi \eta}{c^2 \rb\tb} j^2 + \frac{\epsilon}{\tb} ~~~,
\end{equation}
\begin{equation}
{\cal A}^A = - \frac{1}{r^2\sin\theta}\frac{\pd}{\pd \theta}
\left(\sin\theta \frac{\pd \EE_{\it r}}{\pd \theta}\right) 
- \frac{1}{r^2 \sin^2\theta}\frac{\pd^2 \EE_{\it r}}{\pd \phi^2} +
\frac{1}{r^2} \frac{\pd}{\pd r} \left[\frac{r}{\sin\theta} \left\{
\frac{\pd}{\pd \theta} \left( \sin\theta \EE_\theta \right)
+ \frac{\pd \EE_\phi}{\pd \phi} \right\} \right] ~~~,
\end{equation}
\begin{equation}
{\cal A}^C = \frac{1}{r \sin\theta} \left\{
\frac{\pd}{\pd \theta} \left( \sin\theta \EE_\phi \right)
- \frac{\pd \EE_\theta}{\pd \phi} \right\} ~~~,
\end{equation}
and ${\bf \EE} = {\bf v} \cross \BB$.

The boundary conditions discussed in \S 2.1, expressed here in spectral space,
require that the boundaries be impenetrable
\begin{equation}
W_{\ell m}(r_{bot},t)=W_{\ell m}(r_{top},t)=0 ~~~,
\end{equation}
and stress-free
\begin{equation}
\frac{\p^2 W_{\ell m}}{\p r^2}(r,t)-\left(\frac{2}{r}
+\frac{d \ln \rb}{dr}\right)\frac{\p W_{\ell m}}{\p r}(r,t) = 0 
\mbox{\hspace{.5in}}(r=r_{bot},r_{top}) ~~~,
\end{equation}
\begin{equation}
\frac{\p Z_{\ell m}}{\p r}(r,t)-\left(\frac{2}{r}+\frac{d \ln \rb}{dr}\right)Z_{\ell m}(r,t) = 0 
\mbox{\hspace{.5in}}(r=r_{bot},r_{top}) ~~~.
\end{equation}
We also fix the entropy gradient at the top and bottom boundaries at the value defined
by the initial reference state by requiring the perturbation entropy gradient
to vanish:
\begin{equation}
\frac{\p S_{\ell m}}{\p r}(r_{bot},t) = \frac{\p S_{\ell m}}{\p r}(r_{top},t) = 0 ~~~.
\end{equation}
The magnetic boundary conditions are chosen such that the interior field is continuous
with an external potential field above and below the computational domain:
\begin{equation}
A_{\ell m}(r_{bot},t) = A_{\ell m}(r_{top},t) = 0 ~~~,
\end{equation}
\begin{eqnarray} 
\frac{\p C_{\ell m}}{\p r}(r_{top},t)+\frac{\ell}{r_{top}}C_{\ell m}(r_{top},t) = 0 ~~~,
\frac{\p C_{\ell m}}{\p r}(r_{bot},t)-\frac{\ell+1}{r_{bot}}C_{\ell m}(r_{bot},t) = 0 ~~~.
\end{eqnarray}
For comparison purposes, we also did several simulations in which the magnetic field 
was required to be radial at the boundaries, corresponding to a highly permeable
external medium (Jackson 1999):
\begin{equation}
\frac{\p C_{\ell m}}{\p r}(r,t)=0 \mbox{ and } A_{\ell m}(r,t) = 0 
\mbox{\hspace{.5in}} (r=r_{bot},r_{top}) ~~~.
\end{equation}

Further details on the numerical algorithm are discussed in Clune et al.\ (1999) 

\pagebreak

\begin{table*}[!ht]
\begin{center}
\caption[]{Parameters for the Four Simulations}
\vspace{0.2cm}
\begin{tabular}{||c||cccc||}
\tableline
\tableline
 Case & H & {\em M1} & {\em M2} & {\em M3} \\
\tableline
\tableline
  $N_r, N_{\theta}, N_{\phi}$ & 64, 256, 512 & 64, 256, 512 & 64, 256, 512 &
128, 512, 1024  \\
  $R_a$ & 8.1 $\times 10^4$ & 8.1 $\times 10^4$ & 8.1 $\times 10^4$ & 8.1 $\times 10^4$ \\
  $P_m$ & - & 2 & 2.5 & 4 \\
  $R_c$ & 0.73 & 0.73 & 0.73 &0.73 \\
  $\eta$ (cm$^2$ s$^{-1}$) & - & $7\times 10^{11}$ & $5.6\times 10^{11}$ & $3.5 \times 10^{11}$ \\ 
  $\tau_{\eta}$ (days) & - & 495 & 620 & 990 \\
  $R_e$  & 136 & 136 & 133 & 121 \\
  $R_m$  & - & 272 & 334 & 486 \\
  $\Lambda$ & - & 1.5$\times 10^{-3}$ & 4.5$\times 10^{-2}$ & 20 \\
  $P_e$ & 20 & 17 & 16 & 15\\
  $R_o$  & 0.15 & 0.12 & 0.12 & 0.11 \\
 \tableline
 \tableline
\end{tabular}
\end{center}

The number of radial, latitudinal and longitudinal mesh points are $N_{r}, N_{\theta}, N_{\phi}$.
All simulations have an inner radius $r_{bot}=5.0 \times 10^{10}$ cm and
an outer radius $r_{top}=6.72 \times 10^{10}$ cm and all quantities listed here are 
evaluated at mid-layer depth.   In all cases, $\nu = 1.4 \times 10^{12}$ and
$\kappa = 1.1 \times 10^{13}$ at mid-depth and the Prandtl number 
$P_r = \nu/\kappa$ = 0.125.  Furthermore, the rotation rate of the coordinate
system $\Omega_0 = 2.6 \times 10^{-6} s^{-1}$ in all cases, yielding a 
Taylor number of $T_a = 4 \Omega_0^2 L^4/\nu^2 = 1.2 \times 10^6$, where
$L = r_{top}-r_{bot}$.  Also listed are the Rayleigh number $R_a=(-\p
\rho/\p S)\Delta S g L^3/\rho \nu \kappa$, the magnetic Prandtl number 
$P_m=\nu/\eta$, the convective Rossby number $R_c=\sqrt{R_a/T_a P_r}$, the Reynolds number
$R_e=\vvr' L/\nu$, the magnetic Reynolds number $\tilde{R}_m=\vvr' L/\eta$, the 
Elssasser number $\Lambda=\tilde{B}^2/4\pi\rho\eta\Omega_0$, the P\'eclet number
$P_e=R_eP_r=\vvr' L/\kappa$, the Rossby number 
$R_o=\tilde{v}^\prime/2\Omega_0 L$,
and the ohmic diffusion time $\tau_{\eta}=L^2/(\pi^2 \eta)$,
where $\vvr'$ is the rms convective velocity and $\tilde{B}$ is the rms
magnetic field. A Reynolds number based on the peak velocity at mid depth 
would be about a factor of 5 larger. 

\end{table*}

\pagebreak

\begin{table*}[!ht]
\begin{center}
\caption[]{Representative Velocities, Magnetic Fields, Energies, and Differential Rotation}
\vspace{0.2cm}
\begin{tabular}{||c||cccc||}
\tableline
\tableline
 Case & H & {\em M1} & {\em M2} & {\em M3} \\
 \tableline
 \tableline
 \multicolumn{5}{||c||}{Mid Convective Zone} \\
 \tableline
 $\vrr$  & 61 & 60 & 59 & 58 \\
 $\vtr$  & 63 & 62 & 61 & 54 \\
 $\vphr$ & 137 & 137 & 136 & 104 \\
 $\vphr'$ & 68 & 68 & 67 & 59 \\
 $\vvr$ & 163 & 162 & 161 & 131 \\
 $\vvr'$ & 111 & 111 & 109 & 99 \\
 $\brr$ & - & 21 & 100 & 1752 \\
 $\btr$ & - & 23 & 110 & 1855 \\
 $\bphr$ & - & 29 & 144  & 2277 \\
 $\bphr'$ & - & 28 & 141 & 2239 \\
 $\bbr$ & - & 42 & 207 & 3420\\
 $\bbr'$ & - & 41 & 205 & 3386 \\
 \tableline
 \multicolumn{5}{||c||}{Volume Average} \\ 
 \tableline 
 KE & 9.01$\times 10^6$ & 8.74$\times 10^6$ & 8.96$\times 10^6$  & 5.26$\times 10^6$ \\
 DRKE/KE & 59.3\% & 57.4\% & 57.8\% & 49.5\% \\
 MCKE/KE & 0.3\% & 0.4\% & 0.4\% & 0.5\%\\
 CKE/KE  & 40.4\% & 42.2\% & 41.8\% & 50.0\%\\
 ME & - & $<10^2$ & 1223 & 3.47$\times 10^5$ \\
 ME/KE & - & $<10^{-3}$\% & 0.014\% & 6.6\% \\
 MTE/ME & - & - & 1.4\% & 1.5\%\\
 MPE/ME & - & - & 0.6\% & 0.5\%\\
 FME/ME & - & - & 98\% & 98\% \\
 \tableline
 $\Delta\Omega/\Omega_o$  & 34\% & 34\% & 34\% & 24\% \\
 \tableline
 \tableline
\end{tabular}
\end{center}
Listed for each simulation are the rms amplitude of the velocity
$\vvr$ and each of its components, $\vrr$, $\vtr$, and $\vphr$,
averaged over time at a layer in the middle of the convection
zone. Also listed are the rms amplitudes of the fluctuating total and
zonal velocity, $\vvr'$, and $\vphr'$, obtained after subtracting out
the temporal and azimuthal mean.  For the magnetic simulations, we
include the corresponding rms amplitudes of the magnetic field and its
components, $\bbr$, $\brr$, $\btr$, $\bphr$, $\bbr'$, and $\bphr'$.
Velocities are expressed in m s$^{-1}$ and magnetic fields in $G$.
The kinetic energy density KE ($1/2 ~ \rb v^2$), averaged over volume
and time, is also listed along with the relative contributions from
the non-axisymmetric convection (CKE) as well as the axisymmetric
differential rotation (DRKE) and meridional circulation (MCKE).  We
also list, where appropriate, the average magnetic energy density ME
($B^2/8\pi$) and the relative contribution from each of its
components, including the fluctuating (non-axisymmetric) field FME and
the mean (axisymmetric) toroidal and poloidal fields MTE and MPE.  The
relative latitudinal contrast of angular velocity
$\Delta\Omega/\Omega_o$ between latitudes of $0^{\circ}$ and
$60^{\circ}$ near the top of the domain is also stated for each case
(averaged over both hemispheres).
\end{table*}

\end{document}